# Is it Possible to Extract Metabolic Pathway Information from in vivo H Nuclear Magnetic Resonance Spectroscopy Data?

Alejandro Chinea

Departamento de Física Fundamental, Facultad de Ciencias UNED,
Paseo Senda del Rey nº9, 28040-Madrid - Spain

**Abstract.** In vivo H nuclear magnetic resonance (NMR) spectroscopy is an important tool for performing non-invasive quantitative assessments of brain tumour glucose metabolism. Brain tumours are considered fast-growth tumours because of their high rate of proliferation. In addition, tumour cells exhibit profound genetic, biochemical and histological differences with respect to the original non-transformed cell types. Therefore, there is strong interest from the clinical investigator´s point of view in understanding the role of brain metabolites under normal and pathological conditions and especially in the development of early tumour detection techniques. Unfortunately, current diagnosis techniques ignore the dynamic aspects of these signals. It is largely believed that temporal variations of NMR Spectra are simply due to noise or do not carry enough information to be exploited by any reliable diagnosis procedure. Thus, current diagnosis procedures are mainly based on empirical observations extracted from single averaged spectra. In this paper, firstly a machine learning framework for the analysis of NMR spectroscopy signals which can exploit both static and dynamic aspects of these signals is introduced. Secondly, the dynamics of the signals are further analyzed using elements from chaos theory in order to understand their underlying structure. Furthermore, we show that they exhibit rich chaotic dynamics suggesting the encoding of metabolic pathway information.

## 1 Introduction

The last decade has seen a rise in the application of proton NMR spectroscopy techniques, fundamentally in fields such as biological research [1][2] and clinical diagnosis [3-5]. The main goal within the biological research field is to achieve a deep understanding of metabolic processes that may lead to advances in many areas including clinical diagnosis, functional genomics, therapeutics and toxicology. In addition, metabolic profiles from proton NMR spectroscopy are inherently complex and information-rich, thereby having the potential to provide fundamental insights into the molecular mechanisms underlying health and disease. Nevertheless, it is important to note that the main difficulty is not simply how to extract the information efficiently and reliably but how to do so in a way which is interpretable to people with different technical backgrounds. In fact, machine learning techniques [6] have recently been recognized by biological researchers [7] as an important method for extracting useful information from empirical data.

From a clinical diagnosis point of view, proton NMR spectroscopy has proven to be a valuable tool which has benefited from the knowledge and experience acquired through biological research studies. Furthermore, a large number of proton NMR spectroscopy applications have targeted the human brain. Specifically, it has been extensively used for the study of brain diseases and disorders, including epilepsy [8,9], schizophrenia [10,11], parkinson's disease [12,13] and bipolar disorder [14,15] amongst others. This is mainly due to the fact that proton NMR spectroscopy is a non-invasive technique, which is particularly important in this part of the body where clinical surgery or biopsy is more delicate than in other areas. In addition, it is important to note that it allows in vivo quantification of metabolite concentrations in brain tissue for clinical diagnosis purposes. Moreover, one of the most successful applications of proton NMR spectroscopy has been cancer research [16-19]. In this paper, the focus is on proton NMR brain spectroscopy.

Generally speaking, the process of clinical diagnosis involves the analysis of spectroscopy signals obtained from a well-defined cubic volume of interest (single voxel experiment) in a specific region of the brain during a pre-defined time frame (acquisition time). Two acquisition methods are commonly used, namely point resolved spectroscopy (PRESS) [20] or stimulated echo acquisition mode (STEAM) [21], although the former has gained more popularity because of its ability to provide high signal yield from uncoupled spins. In addition, most of the time, the analysis of the signals is carried out in the frequency domain. The raw signal from the free induction decays (FIDs) is transformed using the discrete fourier transform. Afterwards, a pre-processing stage (e.g. spectral apodization, zero fill, phase correction operations, etc) is also performed to remove artefacts from the acquisition process, thereby improving the characteristics of the signal. Finally, the resulting spectral signals, whose number is approximately equal to the acquisition time divided by the repetition time of the sequence, are averaged and in most cases used for a preliminary diagnosis which relies on a simple visual analysis of the spectra. Otherwise, the resulting spectra are further processed using spectral fitting analysis techniques [22,23] with the goal of determining metabolite concentrations [24-26] for diagnosis purposes [27-28].

Bearing in mind the considerations stated above, the clinical researcher has a strong interest in understanding the role of metabolites under normal and pathological conditions, which is not possible without a deep knowledge of the interactions between them. These interactions are based on a wide set of chemical reactions which are organized into metabolic pathways. Specifically, the chemical reactions involve the transformation of one metabolite into another through a series of steps catalysed by a sequence of enzymes. Surprisingly, the most relevant characteristic of metabolic pathways [29-31] is their universality, i.e. the fact that metabolic pathways are very similar even across quite different species. It is noteworthy that the concept of metabolic pathways is inherently associated with a dynamic context [32,33].

Therefore, extracting information about metabolic pathways from complex biological data sets like those generated through NMR spectroscopy is a challenging research task that requires consideration of the dynamic aspects of these signals.
However, current diagnosis techniques based on proton NMR spectroscopy are still in their infancy. Firstly, as stated above, powerful tools like machine learning techniques are scarcely applied within this context [34]. Indeed, most of the applications of machine learning techniques have been in the field of systems biology research [35-40] and have used data from other techniques like liquid chromatography mass spectrometry or gene expression microarray. This is mainly due to the abovementioned problems regarding the interpretability of the information but also because of a lack of effective communication of results between researchers working in different fields. It is also important to note that current diagnosis techniques ignore the dynamic aspects of these signals. It is largely believed that the information content of temporal variations of NMR Spectra is minimal. Thus, current diagnosis procedures are constrained to empirical observations extracted from a single averaged spectrum. In this paper, a machine learning framework for the analysis of NMR spectroscopy signals is introduced which is able to exploit both static and dynamic aspects of those signals. Secondly, the dynamics of the signals are further analyzed using elements from chaos theory in order to try to understand their underlying structure. Furthermore, the signals are shown to be information-rich, as they have chaotic structure and it is suggested that the signals actually encode metabolic pathway information.

The rest of this paper is organized as follows: In the next section, the problems and difficulties associated with the processing of H-NMR signals are presented. In section 3, a formal characterization of NMR-based data is introduced from a machine learning point of view. The reliability of the proposed measures is assessed through careful analysis of the results provided by a specific experimental design. To complement these results, section 4 focuses on the dynamic aspects of NMR spectral signals. Similarly to section 3, the necessary theoretical background is introduced, followed by the analysis of results provided by a case study designed as a basis for the assessment of hypotheses concerning the structural information of NMR dynamic data. The most important results of the present study are sketched in this section. Finally, section 6 provides a summary of the present study and some concluding remarks.

## 2 Problems Associated with H-NMR Data

Generally speaking, the spectral signals associated with brain metabolites are characterized by one or more peaks at certain resonance frequencies. Furthermore, the molecular structure of a particular metabolite is reflected by a typical peak pattern. In addition, it is also important to note that the area (amplitude) of a peak is proportional to the number of nuclei that contribute to it and therefore to the concentration of the metabolite to which the nuclei belong. Indeed, even if peak amplitudes change from different samples reflecting a change in concentration, the ratios between the central resonance peak and sub-peaks, composing the metabolite fingerprint, always remain constant. Most of the metabolites have multiple resonances many of which are split into multiplets as a result of homonuclear proton scalar coupling. This fact is particularly true of proton NMR spectroscopy at clinical field strengths (from 1.5 up to 3 Teslas), where the whole spectrum occupies a narrow frequency range, normally from -0.8 up to 4.3 parts per million (ppm hereafter) in brain NMR spectroscopy, resulting in significant overlap of peaks from different metabolites. In addition, the response of coupled spins is strongly affected by the acquisition parameters of the NMR sequence (e.g. radio frequency pulses employed and the time intervals set between them [41]). Furthermore, additional difficulties are caused by the presence of uncharacterized resonances from macromolecules or lipids. Specifically, in a typical NMR profile a large number of the resonances may be unassigned, particularly for low level or partially resolved signals. This is further complicated by small but significant sample to sample variations in the chemical shift position of signals, produced by effects such as differences in pH and ionic strength. This kind of positional noise remains a problem as information coming from a given metabolite contaminates the spectral dimensions containing information from other metabolites. In other words, due to this phenomenon the resonance frequency of certain metabolites can suffer slight variations from sample to sample.

In order to address these problems, some sophisticated strategies have been proposed [42-46]. The most popular method of choice has been to use a short total echo time PRESS sequence to optimize signal losses due to scalar coupling evolution [47], and then apply a spectral fitting analysis technique [48-50] to determine metabolite concentrations. The LCModel [50] is the most popular spectral fitting technique. It uses a linear combination of model basis spectra acquired from in-vitro solutions of individual metabolites, including simulated spectra for lipids and macromolecules. It is important to bear in mind that the clinical investigator is interested in obtaining quantitative information about in vivo concentrations of metabolites, the main reason being that it serves as a basis for comparison between pathological and healthy brain tissue. As stated in [51], another strategy consists of optimizing the PRESS echo times so that the target resonance of the metabolite of interest is maintained, while background undesirable peaks are minimized [52,53]. Moreover, in some situations, isolating the desired peak is achieved by using more sophisticated spectral editing techniques like MEGA-PRESS [54] or Multiple Quantum Filters [55,56].

However, the common factor to all the abovementioned techniques is that they completely ignore the inherent dynamics of NMR spectroscopy signals. Unfortunately, it is largely believed that temporal variations of NMR Spectra are simply noise or do not carry enough information to be exploited by any reliable diagnosis procedure. For example, in a single-voxel MRS experiment, as a result of the acquisition process, a whole matrix of data is obtained from the region of interest. Furthermore, within that matrix two consecutive rows correspond to signal frames taken with a time difference equal to the repetition time set for the acquisition sequence (usually 1035-1070 ms for short total echo time sequences). Therefore, the number of rows (signal frames) is approximately equal to the acquisition time divided by the repetition time of the sequence. Indeed, there is a small number of frames that are used for water referencing (usually eight) which are suppressed when generating the data matrix. In addition, each row represents a spectral signal obtained from the volume of interest after a pre-processing stage which transforms the raw FID temporal data into the frequency domain. The frequency domain is usually preferred over the temporal domain [57] since this enables visual interpretation. Moreover, in the frequency domain the NMR signal is represented as a function of resonance frequency. Additionally, each column represents a given metabolite or metabolic signal. The number of columns depends on the particular pre-processing technique used, but is usually 8192 or 4096 dimensions, depending whether or not a zero filling procedure is applied to the transformed signal. In both cases, the associated chemical shift range corresponds approximately to the interval [-14.8, 24.8] ppm. However, as stated above, the range of interest usually taken for brain metabolites is restricted to the interval [-0.8, 4.3 ppm] leading to a dimensionality reduction with respect to the original row size. The exact dimension of these vectors depends on the procedure used for generating the chemical shift scale. The water peak and the creatine peak are commonly used for this purpose. Nevertheless, it is important to note that the resulting matrix after selecting the appropriate range is still composed of high-dimensional row vectors. Afterwards, the temporal variations of the spectra (i.e. the rows of the matrix of data) are averaged in order to obtain a single vector (single spectrum signal). In such vectors each dimension represents the mean value of a particular metabolic signal. As pointed out previously, this vector is further used for spectral fitting techniques purposes. The goal of these techniques is to obtain the parameters of certain model functions that, in some way, fit the Fourier transformed signal with the best accuracy to finally try to quantify metabolite concentrations. The fit metric, the model functions and the method of determining parameters are all choices that the clinical investigator must make. The fit metric commonly used is the chi-squared statistic. Similarly, complex Lorentzian lines or real Gaussian lines are normally used as model functions although more sophisticated approaches also exist [58-60].

## 3 Machine Learning Analysis

This section focuses on the characterization of H-NMR spectroscopy signals from a machine learning point of view. Specifically, we firstly review two pre-processing techniques suited to the main characteristics of NMR-based data, namely the different scales associated with the features (dimensions) of the spectral signals and the existence of correlations between them. Additionally, another feature of NMR data to be considered is its high dimensionality. To this end we also review principal component analysis as a classic tool for dimensionality reduction. In particular, this pre-processing technique will be used in the experimental settings of section 4.

Secondly, we introduce a set of parameters for the structural characterization of NMR-based data sets. The principal advantage of these characterization parameters is that they do not make any assumption about the underlying nature of the data. Therefore, they are appropriate for dealing with both static and dynamic data sets. Finally, some experimental results are also covered which illustrate the application of the proposed characterization scheme to NMR-based data.

### 3.1 Pre-processing Techniques

The goal of pre-processing is to transform the data into a form suited to analysis by statistical procedures. Here, we consider that the FIDs (Free Induction Decays) have already been Fourier Transformed to the frequency domain, and the artefact removal stage carried out. Furthermore, we suppose the data is arranged into a matrix in which each row corresponds to a spectral sample, where two consecutive rows correspond to signal frames taken with a time difference equal to the repetition time of the NMR sequence and each column to a metabolic signal. The metabolic signal corresponds to the spectral intensity at a particular chemical shift.

To determine which is the most suitable pre-processing technique for a given data set, it is important to understand the underlying structure of the input samples. In addition, it is important to note that the metabolic signals (columns in the data matrix) have values that differ significantly, even by several orders of magnitude. Additionally, there are correlations between them (sample dimensions) due to the spectral overlap caused by the proton homonuclear scalar coupling. The choice of pre-processing steps can often have a significant effect on the final performance of a machine learning model. Therefore, it would be desirable to use a pre-processing technique which takes into account the differences in magnitude of metabolic signals, but allowing at the same time the possibility to exploit the existing correlations between them.

#### 3.1.1 Centering and Reduction

This linear rescaling operation [61,62] arranges all the input dimensions to have similar values. Specifically, each input variable is scaled to zero mean and unit variance. This process is recommended when the original features have very different scales as occurs with the metabolic signals of NMR-based data. To do this, we treat each of the input variables (i.e. metabolic signals) independently and for each variable $x_i$ we calculate its mean $\bar{x}_i$ and variance $\sigma_i^2$ with respect to the data set using the following expressions:

$$\bar{x}_i = \frac{1}{N}\sum_{n=1}^{N} x_i^n$$
$$\sigma_i^2 = \frac{1}{N-1}\sum_{n=1}^{N}(x_i^n - \bar{x}_i) \qquad (1)$$

This pre-processing technique will be used in some of the parametric measures to be introduced in the next subsections. The main disadvantage of this normalization procedure is that it treats the input dimensions as independent.

#### 3.1.2 Whitening

The whitening transform [62,63] performs a more sophisticated linear rescaling which allows for correlations amongst the variables. Let us suppose we group the input variables $x_i$ into a vector $\vec{x} = (x_1, x_2, ..., x_d)^t$ , which has sample mean vector and covariance matrix with respect to the N data points of the data set given by:

$$\bar{\xi} = \frac{1}{N}\sum_{n=1}^{N}\vec{x}^n$$
$$\Sigma = \frac{1}{N-1}\sum_{n=1}^{N}(\vec{x}^n - \bar{\xi})(\vec{x}^n - \bar{\xi})^t \tag{2}$$

If we introduce the eigenvalue equation for the covariance matrix, $\Sigma u_j = \lambda_j u_j$ , we can define a vector of linearly transformed input variables given by the following expression:

$$\tilde{x}^n = \Lambda^{-1/2} U^t (\vec{x}^n - \bar{x}) \tag{3}$$

where we have defined:

$$U = (u_1, u_2, ..., u_d)$$
$$\Lambda = diag(\lambda_1, \lambda_2, ..., \lambda_d) \tag{4}$$

It is then straightforward to verify that, in the transformed coordinates, the data set has zero mean and a covariance matrix which is given by the unit matrix.

### 3.1.3 Principal Component Analysis

The principal component analysis (PCA hereafter) is a classic method in machine learning [61,62,64,65,66]. PCA reduces the patterns dimension in a linear way for the best representation in lower dimensions while keeping maximum inertia. The idea is to introduce a new set of orthonormal basis vectors in embedding space such that projections onto a given number of these directions preserve the maximal fraction of the variance of the original vectors. In other words, the error in making the projection is minimized for a given number of directions. The desired principal directions can be obtained as the eigenvectors of the covariance matrix that correspond to the largest eigenvalues. Specifically, features are selected according to the percentage of initial inertia which is covered by the different axes and the number of features is determined according to the percentage of initial inertia to keep for the classification process or any other purpose. When quasi-linear correlations exist between some initial features, these redundant dimensions are removed by PCA. Finally, it is important to note that the best axis for the representation is not necessarily the best axis for discrimination.

### 3.2 Theoretical Background

In the following sub-sections we introduce a set of measures that have been used within the context of supervised learning [66,67]. In this kind of machine learning paradigm knowledge about the problem is represented by means of input-output examples, specifically, examples in the form of vector of attribute values and known classes. The idea is to infer a theory that can predict the classes of unseen cases coming from the same problem.

#### 3.2.1 Inertia

Inertia [62] is a classical measure for the variance of high dimensional data. We distinguish here three types of inertia, namely, global inertia, within-class inertia and between-class inertia.

$$I_G = \frac{1}{N}\sum_{i=1}^{N}\left\|\vec{x}^i\right\|^2$$

$$I_{w_i} = \frac{1}{N_i}\sum_{k=1}^{N_i}\left\|\vec{x}^k - \vec{g}_i\right\|^2 \quad \forall \vec{x}^k \mid class(\vec{x}^k) = w_i$$

$$I_W = \frac{1}{N}\sum_{i=1}^{C} N_i I_{w_i}$$

$$I_B = \frac{1}{N}\sum_{i=1}^{C} N_i \left\|\vec{g}_i\right\|^2 \qquad (5)$$

Global inertia $I_G$ is computed over the entire data set. In contrast, within-class inertia $I_W$, is the weighted sum of the inertia computed on each class, where the weighting is the a priori probability of each class. Between-class inertia $I_B$ is computed on the centers of gravity of each class.

#### 3.2.2 Dispersion and Fischer Criterion

Dispersion and the Fisher criterion (FC hereafter) are two measures [62] for the discrimination or separability between classes (categories defined in the input space). Generally speaking, in a supervised classification problem, classification performance depends on the discrimination power of the features, that is to say, the set of input dimensions which compose the patterns of the data set. It is important to note that the dispersion matrix is not symmetric.

$$FC = \frac{I_B}{I_W}$$

$$D_{ij} = \frac{\left\|\vec{g}_i - \vec{g}_j\right\|}{\sqrt{I_{w_i}}} \qquad (6)$$

As can be deduced, the discrimination is better if the Fisher criterion is large. Similarly, if the dispersion measure between two classes is large then these classes are well separated and the between class distance is larger than the mean dispersion of the classes. In order to apply the Fisher criterion and dispersion measures the data set is normally pre-processed using the linear re-scaling proposed in sub-section 3.1.1. If this measure is close to or lower than one, then the classes are overlapping. In addition, it is important to note that a high degree of overlapping between two classes does not necessarily imply significant confusion between them from the classification point of view. For instance, that is the case for multimodal classes.

### 3.2.3 Confusion Matrix

The confusion matrix [62] is a structural parameter of a data set which provides an estimation of the probability for patterns of one class to be attributed to any other or to the original class. Furthermore, it provides a generic measure of classification complexity. Let us denote as *D* the random variable describing the patterns of the data set. Supposing *D* is a discrete variable, the confusion matrix can be defined as shown below, where *f* is a discriminating function:

$$C_{ij} = \sum_{k \in D} p(D_k \mid w_i) f_j(D_k) \tag{7}$$

More specifically, a classifier is always defined in terms of its discriminating function *f* which divides the n-dimensional input space into as many regions as there are classes. If there are C classes $w_i$ $1 \leq i \leq C$, the discriminant function *f* may also be expressed in terms of the following indicator function $f_i$, where $f_i(u) = 1$ if $f(u) = w_i$ and $f_i(u) = 0$ otherwise. The classifier performance may also be expressed by the averaged classification error

$$E(f) = \sum_{i=1}^{C} p_i \sum_{j \neq i} C_{ij}(f) \tag{8}$$

The best confusion matrix is that corresponding to the Bayesian classifier (minimal attainable classification error). It can be deduced that the confusion matrix cannot be computed using the Bayesian classifier as it would imply a perfect knowledge of the statistics of the problem (conditional probabilities $p(D|w_i)$ and the a priori probabilities $p_i$). Therefore, the best confusion matrix must be in practice approximated. To this end, the *k*-nearest neighbour classifier [61,63] is often used because of its powerful probability density estimation properties. More specifically, a set of values for *k* are generated, for instance the following odd sequence *k*=1,3,5,7,9,11. Afterwards, a leave-one-out cross-validation procedure [63,66,68] is performed for the entire dataset for each *k* from the selected set of values. Finally, the best confusion matrix is that obtained for the value of *k* which minimizes the performance error defined in expression (8).

### 3.2.4 Fractal Dimension

The fractal dimension [69,70] is a parameter which measures the intrinsic dimension or degrees of freedom of a data set. Furthermore, it can be defined as the dimension of the sub-manifold structure of the data. Generally speaking, a classifier is designed for input patterns with a user-defined dimension, let us say d. This samples distribution can have intrinsically less than d degrees of freedom. Therefore, knowledge about this local dimension is useful as information not only because it can better represent the complex nature of real objects but also because it can be conveniently used by pre-processing methods [62,64] to simplify the problem at hand by means of a dimensionality reduction of the data set. However, it is important to note that the choice of a suitable pre-processing technique is not obvious as the adequate projection can be a non-linear one [71-74].

An important variety of definitions of the fractal dimension has been proposed [75]. Here we consider the similarity dimension (see expression (9)), that is, how a data set object remains statistically similar independently of the scale of observation. For instance, what is seen as a point at a very large scale, could appear as a sphere (three-dimensional object) at some smaller scale or as a strongly intertwined line (one dimensional object).

$$d = \lim_{r \to 0} \frac{\log(N(r))}{\log(1/r)} \quad (9)$$

The computation of the fractal dimension of an object can be performed in different ways. The most popular method is the box counting method [75]. Basically, this method consists in counting the number of hypercubes *N(r)* that contain samples (patterns) in a space divided in hypercubes of side *r*. However, as stated in [62], this method is not suitable to obtain the fractal dimension if *r* is too large, since the dimension value tends to zero. Similarly, real data sets are not continuous. The clouds of samples that composes the data sets become isolated as *r* tends to zero, and the fractal dimension for too small a value of *r* becomes zero again. Furthermore, the noise dimension may also overlap the true data dimension for small *r*. Therefore the fractal dimension has to be found for intermediate values of *r*, and in some cases the exact value is rather difficult to obtain for instance because the number of samples may be insufficient or the data is noisy and changes the estimation of the intrinsic dimension. In order to provide an aid to the decision, the curve of *log(N(r))* against *log(1/r)* of the data set is plotted. Then, from the observation of the curve slopes a reasonable guess of the fractal dimension can be obtained.

To compute the similarity dimension of equation (9), here we consider a more robust algorithm [76]. The basic idea is to generate a unique code for each elementary hypercube, and then count the number of different codes generated by the distribution samples. The algorithm may be described by the following steps:

1. All the distribution points are set non-negative by translation. The initial hypercube side $r_0$ that completely covers the data set is calculated.

2. A new $r$ is obtained in a logarithmic scale.

3. For each sample vector $\vec{x}^i$, a code $\pi^i = \{\pi_1^i, \pi_2^i, ..., \pi_d^i\}$ is built as the integer division of each component by $r$:

$$\pi_k^i = \left\lfloor \frac{x_k^i}{r} \right\rfloor$$

4. The number $N(r)$ of different $\pi^i$ codes is counted.

5. For $N(r) < M/2$ go to step 2. Since $M$ is the number of samples, it seems a reasonable condition to stop the algorithm when a mean value of 2 samples is inside a hypercube.

6. Finally, a linear regression is performed to obtain the mean slope of the log-log curve of the pairs $(r, N(r))$.

### 3.3 Data Set Description

For experimental purposes two data sets were used both of them corresponding to short-TE NMR single voxel brain spectroscopy. Let us denote the first data set as A, which corresponds to data collected from 11 healthy patients of ages ranging from 25 up to 45, with a mean of 31.45 years. The data was collected from different brain regions (see table 1 for details) and with approximately equal voxel sizes. The acquisition time was approximately equal to 5 minutes for all patients, using a total echo time (TE) equal to 23ms and a repetition time (TR) of 1070ms.

| Data Set A<br><br>(TE= 23ms , TR= 1070ms) | Voxel Location | Voxel Size |
|---|---|---|
| [$A_1$] | [L,P,I] = [0.9, 6.3, 17.2] | [20, 20 , 20] |
| [$A_2$] | [L,P,S] = [8.1, 27.7, 51.5] | [28.7, 20, 26.7] |
| [$A_3$] | [L,P,S] = [6.7, 9.5, 15.1] | [20, 20 , 20] |
| [$A_4$] | [L,P,S] = [0.3, 6.6, 44.9] | [20, 20 , 20] |
| [$A_5$] | [L,P,S] = [1.2, 18.5, 61.6] | [20, 20 , 20] |
| [$A_6$] | [R,P,S] = [0.3, 14.6, 68.8] | [20, 20 , 20] |
| [$A_7$] | [L,P,S] = [2.79,25.97,60.89] | [20, 18.24, 15.4] |
| [$A_8$] | [R,P,S] = [21.3, 92.8, 43.2] | [20, 20 , 20] |
| [$A_9$] | [L,P,S] = [27.4, 25, 63.8] | [20, 20 , 20] |
| [$A_{10}$] | [L,P,S] = [23.2, 26.1, 38.7] | [20, 20 , 20] |
| [$A_{11}$] | [R,P,S] = [4.1, 13.3, 45.1] | [28.7, 20, 26.8] |

Table 1. Data set A

Similarly, the second data set (see table 2 for details) is a small data set used for comparison purposes in the experimental settings of section 3 and it is composed of data also collected at short TE but with a slightly different parameterization. Let us denote the second data set as B. For these data, the total echo time was set to 35ms and the repetition time to 1500ms and the patients´ ages ranged from 30 up to 45 with a mean of 36 years. In addition, the data corresponds to three patients, where data matrix $B_1$ and $B_3$ belong to two healthy patients, while data matrix $B_2$ corresponds to a patient that was diagnosed with a tumour (after a rigorous clinical diagnosis procedure including biopsy). In particular, the data matrix corresponding to patient $B_2$ represents data obtained exactly from the brain tumour area.

| Data Set B<br><br>(TE= 35ms, TR= 1500ms) | Voxel Location | Voxel Size |
|---|---|---|
| [$B_1$] | [L,A,S] = [19.4, 14.8, 96.3] | [16.5, 17.7 , 17] |
| [$B_2$] | [R,A,S] = [20.9, 35.5, 76.2] | [20, 29.6, 20] |
| [$B_3$] | [L,A,S] = [30, 16, 64.7] | [20, 20 , 20] |

Table 2. Data set B

All the spectral data were generated and pre-processed using a spectroscopic and processing software package from GE Medical Systems (documentation is available at http://www.wpic.pitt.edu/research/psychosis-insight/links/SAGELXGuide.pdf). This tool comes with a set of built-in functions (macro reconstruction operations) which provide different useful processing options of raw FID data. We used a macro reconstruction operation which provides internal water referencing, spectral apodization, zero filling, convolution filtering and Fourier transform operation on each of the acquired frames. However, it is important to note that the convolution filtering and water suppression options were not selected. The result of this processing step is a data matrix where each column represents a temporal series spectrum of a specific metabolic signal and each row represents a sample or pattern from the brain region of interest. Each sample belongs to a space of approximately 1064 dimensions corresponding to a chemical shift range of [-0.8, 4.3] ppm. As mentioned in section 2, this interval corresponds to the range where the main resonances concerning the 35 known metabolites involved in brain metabolism are located.

### 3.3 Experimental Design (Case Study I)

In this section all the data described in the previous section and used for the experiments of the current section was centered and reduced for normalization. It is also important to note that although not reported here, similar results are obtained when using the whitening transformation.

The first experiment conducted was designed to check the variance of the measures corresponding to different healthy subjects. Furthermore, the idea was also to assess the possible existence of outliers. At this point, it is important to remember that both data sets described in the previous section are dynamic. In addition, the set of parametric measures introduced in section 3.2 were proposed in a supervised learning context. This means that the samples of the data set must be rated as belonging to a predefined set of categories. In our case, the categories are defined according to the number of different subjects that compose the database. For data set A there are samples coming from eleven different individuals, therefore according to the proposed schema we have eleven different categories for the samples. Moreover, samples belonging to subject $A_i$ are rated as belonging to the class $C_i$ where $i=1,2,\ldots 11$. At this point, it is important to highlight the fact that we have chosen this categorization scheme for two reasons: firstly, as stated above, in order to check the performance of the proposed structural parameters and secondly, because of a lack of data from patients presenting disorders that could bias the results. Ideally, for diagnosis purposes we would have used just two categories for discriminating disease.

| Dispersion Matrix (data set A) | $A_1$ | $A_2$ | $A_3$ | $A_4$ | $A_5$ | $A_6$ | $A_7$ | $A_8$ | $A_9$ | $A_{10}$ | $A_{11}$ |
|---|---|---|---|---|---|---|---|---|---|---|---|
| $A_1$ | 0 | 0.7607 | 0.2124 | 0.2490 | 0.1085 | 0.1496 | 0.3068 | 0.2789 | 0.2014 | 0.2724 | 0.9519 |
| $A_2$ | 0.7998 | 0 | 0.8059 | 0.6706 | 0.7335 | 0.7309 | 1.1255 | 0.9680 | 0.7464 | 0.7792 | 0.4634 |
| $A_3$ | 0.2233 | 0.8061 | 0 | 0.2984 | 0.1700 | 0.2188 | 0.2844 | 0.2687 | 0.2370 | 0.3028 | 1.0238 |
| $A_4$ | 0.2660 | 0.6815 | 0.3032 | 0 | 0.2400 | 0.2376 | 0.4631 | 0.3939 | 0.2665 | 0.2788 | 0.8375 |
| $A_5$ | 0.1207 | 0.7759 | 0.1797 | 0.2499 | 0 | 0.1289 | 0.2571 | 0.2380 | 0.1926 | 0.2445 | 0.9742 |
| $A_6$ | 0.1555 | 0.7222 | 0.2162 | 0.2311 | 0.1205 | 0 | 0.3158 | 0.2850 | 0.1811 | 0.2186 | 0.9152 |
| $A_7$ | 0.2655 | 0.9265 | 0.2340 | 0.3751 | 0.2001 | 0.2631 | 0 | 0.1651 | 0.3118 | 0.3334 | 1.1622 |
| $A_8$ | 0.2913 | 0.9616 | 0.2669 | 0.3851 | 0.2235 | 0.2864 | 0.1993 | 0 | 0.3295 | 0.3428 | 1.1842 |
| $A_9$ | 0.2035 | 0.7173 | 0.2277 | 0.2520 | 0.1750 | 0.1761 | 0.3641 | 0.3188 | 0 | 0.2155 | 0.8839 |
| $A_{10}$ | 0.2625 | 0.7142 | 0.2775 | 0.2514 | 0.2118 | 0.2028 | 0.3712 | 0.3163 | 0.2056 | 0 | 0.8796 |
| $A_{11}$ | 0.8310 | 0.3848 | 0.8499 | 0.6843 | 0.7647 | 0.7690 | 1.1724 | 0.9899 | 0.7637 | 0.7969 | 0 |

Table 3. Dispersion matrix for data set A.

Table 3, shows the results obtained after computing the dispersion for data set A after the categorization procedure described above. The first thing to note is that most of the dispersion values are below one, thereby indicating a high degree of overlapping between classes. Furthermore, these results indicate the absence of outliers during the measurement process as their existence would have led to significant differences between the dispersion values. In addition, the values obtained are logical taking into account that all the samples were collected under the same settings (same repetition time and echo time). However, it is important to note that the dispersion values associated with categories $C_2$ and $C_{11}$ with respect to the rest of the categories (rows and columns $A_2$ and $A_{11}$ of the data matrix) are slightly higher when compared with the rest of the elements of the matrix. Indeed, there are dispersion values which are close to unity or even slightly higher than unity. A careful analysis revealed that this effect was caused by the voxel size. The size of the voxel for the “x” and “z“ dimensions (see table 1) is slightly bigger for categories $C_2$ and $C_{11}$ with respect to the standard voxel size [20,20,20]. This is also true for dimensions “y” and “z” on the voxel associated with class $C_7$. We observed that the effect is to some extent proportional to the discrepancy between the actual size and the standard voxel size.

In order to further validate the results obtained with the dispersion matrix we computed the Fisher criterion obtaining a value of 0.8504 which confirmed the expected overlapping between classes.

| Confusion Matrix (data set A) | $A_1$ | $A_2$ | $A_3$ | $A_4$ | $A_5$ | $A_6$ | $A_7$ | $A_8$ | $A_9$ | $A_{10}$ | $A_{11}$ |
|---|---|---|---|---|---|---|---|---|---|---|---|
| $A_1$ | 100 | 0 | 0 | 0 | 0 | 0 | 0 | 0 | 0 | 0 | 0 |
| $A_2$ | 0 | 99.4505 | 0 | 0 | 0 | 0 | 0 | 0 | 0 | 0 | 0.5495 |
| $A_3$ | 0 | 0 | 100 | 0 | 0 | 0 | 0 | 0 | 0 | 0 | 0 |
| $A_4$ | 0.5263 | 0 | 0 | 98.9474 | 0 | 0 | 0 | 0 | 0.5263 | 0 | 0 |
| $A_5$ | 8.4211 | 0 | 0 | 0.5263 | 91.0526 | 0 | 0 | 0 | 0 | 0 | 0 |
| $A_6$ | 14.7368 | 0.5263 | 0.5263 | 1.0526 | 1.5789 | 79.4737 | 0 | 0 | 1.0526 | 1.0526 | 0 |
| $A_7$ | 0 | 0 | 0 | 0 | 0 | 0 | 100 | 0 | 0 | 0 | 0 |
| $A_8$ | 0 | 0 | 0 | 0 | 0 | 0 | 1.5789 | 98.4211 | 0 | 0 | 0 |
| $A_9$ | 0 | 0 | 0 | 0 | 0 | 0 | 0 | 0 | 100 | 0 | 0 |
| $A_{10}$ | 0 | 0 | 0 | 0 | 0 | 0 | 0.5263 | 0 | 0 | 99.4737 | 0 |
| $A_{11}$ | 0 | 0 | 0 | 0 | 0 | 0 | 0 | 0 | 0 | 0 | 100 |

Table 4. Confusion matrix for data set A

In addition, following a similar procedure we computed the confusion matrix associated with the eleven categories composing data set A. Table 4 shows the results of the computation. It is important to note that the estimations of conditional probabilities between classes shown in the table are multiplied by a factor of 100 to get percentage values. Therefore, it is easy to deduce that there is no apparent confusion between categories as most of the values are zero or close to zero. Nevertheless, samples belonging to class $C_6$ are apparently the most difficult to classify. Generally speaking, a high degree of overlapping given by the dispersion does not necessarily mean significant confusion between the classes from a classification point of view. This is an indication of the existence of multi-modal or very elongated categories.

The second experiment conducted was designed to check the influence of the parameters of the PRESS sequence while including data samples from a patient who was diagnosed with a tumour, in order to check the suitability of the characterization parameters introduced in section 3.2. To this end, we merged the two available data sets A and B (see section 3.3 for details) to create a unique database. We followed the same categorization scheme explained before consisting of assigning as many categories as the number of patients, where data samples belonging to the same patient were assigned to the same category.

| Dispersion Matrix (data set A+B) | $A_1$ | $A_2$ | $A_3$ | $A_4$ | $A_5$ | $A_6$ | $A_7$ | $A_8$ | $A_9$ | $A_{10}$ | $A_{11}$ | $B_1$ | $B_2$ | $B_3$ |
|---|---|---|---|---|---|---|---|---|---|---|---|---|---|---|
| $A_1$ | 0 | 0.75 | 0.21 | 0.24 | 0.10 | 0.15 | 0.30 | 0.27 | 0.20 | 0.26 | 0.93 | 0.45 | 2.24 | 0.20 |
| $A_2$ | 0.79 | 0 | 0.79 | 0.66 | 0.72 | 0.72 | 1.11 | 0.96 | 0.74 | 0.77 | 0.46 | 1.32 | 3.14 | 0.88 |
| $A_3$ | 0.22 | 0.79 | 0 | 0.29 | 0.17 | 0.22 | 0.28 | 0.27 | 0.23 | 0.30 | 1.00 | 0.42 | 2.37 | 0.23 |
| $A_4$ | 0.26 | 0.67 | 0.30 | 0 | 0.23 | 0.24 | 0.45 | 0.39 | 0.26 | 0.27 | 0.82 | 0.60 | 2.08 | 0.29 |
| $A_5$ | 0.12 | 0.76 | 0.18 | 0.24 | 0 | 0.12 | 0.25 | 0.23 | 0.19 | 0.24 | 0.95 | 0.41 | 2.22 | 0.18 |
| $A_6$ | 0.15 | 0.71 | 0.21 | 0.23 | 0.12 | 0 | 0.31 | 0.28 | 0.18 | 0.21 | 0.89 | 0.47 | 2.05 | 0.21 |
| $A_7$ | 0.26 | 0.91 | 0.23 | 0.37 | 0.19 | 0.26 | 0 | 0.16 | 0.31 | 0.32 | 1.14 | 0.23 | 2.47 | 0.23 |
| $A_8$ | 0.29 | 0.95 | 0.26 | 0.38 | 0.22 | 0.28 | 0.19 | 0 | 0.32 | 0.33 | 1.16 | 0.20 | 2.39 | 0.26 |
| $A_9$ | 0.20 | 0.71 | 0.22 | 0.25 | 0.17 | 0.17 | 0.35 | 0.31 | 0 | 0.21 | 0.86 | 0.51 | 2.05 | 0.28 |
| $A_{10}$ | 0.25 | 0.70 | 0.27 | 0.25 | 0.20 | 0.20 | 0.36 | 0.31 | 0.20 | 0 | 0.86 | 0.51 | 1.67 | 0.29 |
| $A_{11}$ | 0.81 | 0.38 | 0.83 | 0.67 | 0.75 | 0.76 | 1.15 | 0.98 | 0.75 | 0.79 | 0 | 1.35 | 2.89 | 0.93 |
| $B_1$ | 0.38 | 1.06 | 0.33 | 0.48 | 0.31 | 0.38 | 0.23 | 0.16 | 0.42 | 0.45 | 1.30 | 0 | 2.74 | 0.34 |
| $B_2$ | 0.58 | 0.77 | 0.58 | 0.50 | 0.51 | 0.51 | 0.74 | 0.60 | 0.53 | 0.45 | 0.85 | 0.84 | 0 | 0.62 |
| $B_3$ | 0.19 | 0.81 | 0.21 | 0.26 | 0.15 | 0.20 | 0.26 | 0.24 | 0.27 | 0.29 | 1.02 | 0.38 | 2.33 | 0 |

Table 5. Dispersion matrix for data set A+B.

Let us denote the merged data set as A+B. It is important to remember that the samples associated with data set B were collected using a different parameterization sequence from that used for data set A. In particular, for data set B the echo time used was 35ms and the repetition time 1500 ms. The fisher criterion computed for this data set led to a value of 0.8262 indicating overlapping between the defined categories.

Table 5 shows the results of computing the dispersion matrix for the data set A+B. The first thing that can be gleaned from the table is that most of the values are below one, thereby indicating the existence of overlapping between classes from the dispersion point of view. Therefore, from a dispersion point of view there is not too much difference between data samples coming from the two different parameterizations, although this can be considered to some degree a logical result taking into account that the echo time of the two sequences are relatively close. Nevertheless, the samples associated with the class representing a disorder (i.e: a patient with a tumour) led to dispersion values much higher than for the rest of the values found in the table, even taking into account the voxel size effect. More specifically, they are very close or bigger than two (see column $B_2$ from table 5) indicating a dispersion caused by the existence of the disease. At this point, it is important to remember that the dispersion matrix is not symmetric. In addition, these results seem to indicate that classes from healthy patients are well separated from the class indicating a disease. Although not shown here, this was also confirmed by the confusion matrix. Specifically, from a classification point of view there is no apparent confusion between the category representing the patient with a tumour and the other categories.

Finally, despite our previous considerations concerning the amount of data, in order to deepen and further complement the previous results, we conducted an experiment consisting of using the entire database A+B but now defining only two categories $C_0$ and $C_1$ to indicate "health" and "disease" respectively. The categorization procedure is similar to the procedure described above. All the samples belonging to data matrix $B_2$ were rated as belonging to class $C_1$, while the rest of the samples were rated as belonging to category $C_0$. The results of the dispersion matrix and confusion matrix computation are shown in table 6. From the inspection of the table we appreciate that there is a slight confusion for the recognition of category $C_1$ ("disease"). Specifically, there is a certain probability that patterns belonging to the class "health" may be rated as belonging to the class "disease", although this probability is small, and from a diagnosis point of view an error of this kind would be less serious when compared to the opposite case which is apparently inexistent.

| Dispersion Matrix | $C_0$ | $C_1$ |
|---|---|---|
| $C_0$ | 0 | 1.9655 |
| $C_1$ | 0.1249 | 0 |

| Confusion Matrix | $C_0$ | $C_1$ |
|---|---|---|
| $C_0$ | 95.2721 | 0 |
| $C_1$ | 4.7279 | 100 |

Table 6. Dispersion and confusion matrixes for data set A+B when considering only two categories in the input space: "health" and "disease".

To summarize, although further experiments must be carried out, for instance using a much larger amount of data, these preliminary results have shown that machine learning characterization of spectral data has proven to be useful not only for the detection of outliers but also as a starting point for reliable diagnosis systems development based on NMR spectroscopy data.

## 4 Chaos Theory Analysis

This section focuses on the study and characterization of the dynamics associated with NMR data. Many processes in nature present a nonlinear structure, although the possible nonlinear nature might not be evident in specific aspects of their dynamics. Therefore, before applying nonlinear techniques, for example those inspired by chaos theory, it is necessary to first ask if the use of such advanced techniques is justified by the data. Bearing in mind these considerations, in this section we consider the analysis of time series of spectra. Specifically, as opposed to standard single-voxel spectroscopy experiments where the resulting spectral signals are averaged in what follows we consider the entire set of frames obtained after the NMR acquisition process.

After a short introduction to the theory of dynamical systems [77], some basic information-theoretic concepts are presented. These concepts will be used to introduce the most important features of chaos theory [78-81] that will be used, in turn, as a basis for the subsequent analysis. In particular, the dynamic aspects of NMR-based data are studied in detail through a careful analysis formulated in terms of experimental design.

## 4.1 Theoretical Background

A dynamical system is a system whose state varies with time. More specifically, it is composed of two parts, namely, a state and a dynamic. A state describe the current condition of the system, usually as a vector of observable quantities. The dynamic of a system encapsulates the process of change over time. Moreover, time evolution as a property of a dynamical system can be measured by recording time series. The most extended mathematical description for dynamical systems is the state-space model [66,82,83,84]. According to this model, we think in terms of a set of state variables whose values at any particular instant of time are supposed to contain sufficient information to predict the future evolution of the system. Let $x_1(t), x_2(t),\ldots, x_d(t)$ denote the state variables of a nonlinear dynamical system, where $d$ is the order of the system.. The dynamics of a large class of nonlinear dynamical systems may be cast in the form of a system of first-order nonlinear differential equations as follows:

$$\frac{d}{dt}\vec{x}(t) = F(\vec{x}(t)) \qquad \vec{x}(t) = [x_1(t), x_2(t), \ldots\ldots, x_d(t)]^t \tag{10}$$

The sequence of states exhibited by a dynamical system during this evolution is called a trajectory. These trajectories often approach characteristic behaviour in the limit known as an attractor. An attractor is a region in state space from where all trajectories converge from a larger subset of state space. They are manifolds of dimensionality lower than that of the state space. In addition, in the temporal limit, only four types of qualitatively different dynamic regimes can be identified: fixed points, limit cycles, quasi-periodicity and chaos.

Moreover, the manifold may consist of a single point in the state space, in which case we speak of a fixed point. Alternatively, it may be in the form of a periodic trajectory, in which case we speak of a limit cycle. When the limit cycle behaviour is not restricted to a single periodicity then we speak of quasi-periodic behaviour. Finally, when the dynamic repeatedly stretches, compresses and folds the state space, chaotic attractors emerge. The trajectory of a system following a chaotic regime is aperiodic. Chaotic attractors exhibit highly complex behaviour. Their main characteristic is their sensitivity to initial conditions. Furthermore, the most interesting feature is that the system in question is deterministic in the sense that its operation is governed by fixed rules. Nevertheless, such a system with only a few degrees of freedom can exhibit very complex behaviour that looks random. From an information-theoretic point of view, chaotic systems are producers of information. That is, their states can be viewed as reservoirs of information.

#### 4.1.1 Entropy and Mutual Information

Entropy and mutual information [85,86] constitute the most classical information theoretic measures. In order to introduce these concepts, let us consider a random variable X. Each realization of this variable can be viewed as a message sent by the random variable. Supposing that X is a discrete random variable, the entropy of variable X denoted as H(X) is defined as the average amount of information conveyed per message sent by the random variable X.

$$H(X) = \sum_{x \in X} p(x) \log(p(x)) \tag{11}$$

Similarly, mutual information I(X,Y) between random variables X and Y is a measure of the uncertainty about random variable Y that is resolved by observing random variable X. It can be viewed as the amount of information we gain about variable Y by observing variable X. Supposing X and Y are discrete random variables, the mutual information can be expressed as:

$$I(X,Y) = \sum_{x \in X} \sum_{y \in Y} p(x,y) \log\left(\frac{p(x,y)}{p(x)p(y)}\right) = H(X) - H(X \mid Y) \tag{12}$$

As can be deduced, the mutual information is zero when the two random variables under consideration are independent.

#### 4.1.2 Poincaré Plots

Although there are many ways of visualizing the properties (e.g. chaotic, random, etc) of a time series [87-89], the simplest way is by means of a Poincaré plot. The Poincaré plot is a graphic representation of the correlation between consecutive interval series (phase space representation). It represents the time series $\{x(n)\}_{n=1}^{N}$ against a delayed version $\{x(n-\tau)\}_{n=1}^{N}$, where τ is the delay or lag (usually set as τ = 1 ). The geometry of the Poincaré plot is essential for identification of the underlying characteristics of a time series. For instance, the phase space of a random time series leads to a graphic with points uniformly distributed all over the graph. In other words, there is no structure in the graph as the points are uncorrelated. Conversely, the points of the phase space associated with a quasi-periodic or a chaotic time series display a specific structure reflecting the particular characteristics of the underlying dynamical system which generated the data. A common way to describe the geometry of the diagram [90] is to fit an ellipse to the graph. The ellipse is fitted onto the line of identity at 45 degrees to the normal axes. The standard deviation of the data points perpendicular to the line of identity describes short term variability of the stochastic process. Similarly, the standard deviation along the line of identity describes long term variability. These parameters (short-term and long-term standard deviations) are sometimes used as features to feed machine learning models.

#### 4.1.3 Phase Space Reconstruction

Generally speaking, a time series can be viewed as a sequence of observations coming from a dynamical system. Such sequences of observations usually do not properly represent the multidimensional phase space of the dynamical system. Therefore, we need some technique to unfold the multidimensional structure using the available data. Phase space reconstruction may be defined as the identification of mapping that provides a model for an unknown dynamical system. More specifically, given a time series $\{y(n)\}_{n=1}^{n=N}$ we wish to build a model that captures the underlying dynamics responsible for generation of the observable *y(n)*. The most important phase space reconstruction technique is the method of delays [91,92]. According to this method, the dynamics of the unknown system can be unfolded in a *D*-dimensional space constructed from the following vector, where τ is called the embedding delay:

$$y_R(n) = [y(n), y(n-\tau), y(n-2\tau), \ldots\ldots, y(n-(D-1)\tau)] \tag{13}$$

Moreover, given y(n) for varying discrete time *n* of an unknown dynamical system, phase space reconstruction is possible using the *D*-dimensional vectors of expression (13) provided that $D \geq 2d+1$, where *d* is the dimension of the state space of the system. The procedure for finding a suitable *D* is called embedding and the minimum integer *D* that achieves dynamic reconstruction is called the embedding dimension.

#### 4.1.4 Lyapunov Exponents

The Lyapunov exponents [93] are statistical quantities that describe the uncertainty about the future state of an attractor. More specifically, they quantify the exponential rate at which nearby trajectories separate from each other while moving on the attractor. Let *x(0)* be an initial condition and *{x(n), n=0,1,2,.....}* the corresponding trajectory. Consider an infinitesimal displacement from the initial condition *x(0)* in the direction of a vector *y(0)* tangential to the orbit. Then, the evolution of the tangent vector determines the evolution of the infinitesimal displacement of the perturbed trajectory *{y(n), n=0,1,2,…}* from the unperturbed trajectory *x(n)* and the ratio $\frac{\|y(n)\|}{\|y(0)\|}$ is the factor by which the infinitesimal displacement grows or shrinks. For an initial condition x(0) and initial displacement $\alpha_0 = \frac{y(0)}{\|y(0)\|}$ the Lyapunov exponent is defined

by:

$$\lambda(x(0),\alpha) = \lim_{n\to\infty} \frac{1}{n} \log\left( \frac{\|y(n)\|}{\|y(0)\|} \right) \tag{14}$$

Therefore, Lyapunov exponents account for the sensitivity of a chaotic process to initial conditions. It is important to note that a *d*-dimensional chaotic process has a total of *d* Lyapunov exponents. Positive Lyapunov exponents account for the instability of a trajectory throughout the state space. Conversely, negative Lyapunov exponents govern the decay of transients in the trajectory. In addition, the largest Lyapunov exponent also defines the horizon of predictability of a chaotic process [94]. Taking into account the definitions stated above, we can define a chaotic process as a process generated by a nonlinear deterministic system with at least one positive Lyapunov exponent [66]. The maximal Lyapunov exponent [95,96] can be determined without the explicit construction of a model for the time series. A reliable characterization requires that the independence of embedding parameters and the exponential law for the growth of distances are checked explicitly.

### 4.2 Experimental Design (Case Study II)

In this section, we are mainly concerned with the analysis of time series of spectra. This means that the time ordering of the data is supposed to contain a significant part of the information. Therefore, in order to capture information about the unknown underlying dynamics the observations must be equally spaced in time. Accordingly, we used the data set A, which was explained in section 3, after applying the whitening transformation presented in section 3.1. In addition, it is important to note that the information theoretic measures (entropy and mutual information) used in this section for unravelling some aspects of the structure of the data were computed using the software package [97].

On the other hand, for a given nucleus, the amplitude and frequency are the most specific parameters of individual resonances in NMR spectra. In the case of scalar coupled spins, the resonances are split into several smaller resonances according to a well-defined pattern which is dependent on the other coupled spins. A detailed compilation of proton chemical shifts of the groups (i.e. methyl, methylene, amines etc) which compose brain metabolites is presented in table 7. This table was built by decomposing the chemical shift range associated with the 35 brain metabolites presented in [98] into bins of 0.1 ppm resolution width.

From inspection of the table, it can be observed that there are regions which are more populated with molecular group contributions than others. In particular, the interval [3.0, 4.0] supports the biggest amount of group contributions when compared to the rest of the intervals. Furthermore, we can appreciate certain intervals where apparently there is no contribution at all of any of the molecular groups. For instance, the interval [-0.8, 0.9] constitutes a clear example of this. Taking into account these observations, it is reasonable to think that increased complexity should be observed in metabolic signals associated with the intervals where there are more contributions of molecular groups. In addition, it would be desirable to quantify the complexity of any such increment. A natural way of assessing the complexity associated with metabolic signals is to use a complexity measure such as the fractal dimension introduced in section 3.

| | | | | | | | | | |
|---|---|---|---|---|---|---|---|---|---|
| [-0.9,-0.8] | [-0.8,-0.7]] | [-0.7,-0.6]] | [-0.6,-0.5] | [-0.5,-0.4] | [-0.4,-0.3] | [-0.3,-0.2] | [-0.2,-0.1] | [-0.1,0.0] | |
| | | | | | | | | | |
| [0.0,0.1] | [0.1,0.2] | [0.2,0.3]] | [0.3,0.4] | [0.4,0.5] | [0.5,0.6] | [0.6,0.7] | [0.7,0.8] | [0.8,0.9] | [0.9,1.0] |
| | | | | | | | | | CH3 |
| [1.0,1.1] | [1.1,1.2] | [1.2,1.3] | [1.3,1.4] | [1.4,1.5] | [1.5,1.6] | [1.6,1.7] | [1.7,1.8] | [1.8,1.9] | [1.9,2.0] |
| CH3 | | | CH3<br>CH3 | CH3 | | | | CH2<br>CH2 | CH3 |
| [2.0,2.1] | [2.1,2.2] | [2.2,2.3] | [2.3,2.4] | [2.4,2.5] | [2.5,2.6] | [2.6,2.7] | [2.7,2.8] | [2.8,2.9] | [2.9,3.0] |
| CH3<br>CH3 | CH2<br>CH2<br>CH2<br>CH2<br>CH2 | CH2<br>CH | CH2<br>CH2<br>CH2<br>CH2<br>CH2<br>CH2 | CH2<br>CH2<br>CH2 | CH2<br>CH2 | CH2<br>CH2 | | CH2 | CH2<br>CH2<br>CH2<br>CH2<br>CH2 |
| [3.0,3.1] | [3.1,3.2] | [3.2,3.3] | [3.3,3.4] | [3.4,3.5] | [3.5,3.6] | [3.6,3.7] | [3.7,3.8] | [3.8,3.9] | [3.9,4.0] |
| CH2<br>N(CH3)<br>CH2<br>N(CH3)<br>CH2 | N(CH3)3<br>CH2<br>CH2<br>CH2<br>CH2<br>CH2 | N(CH3)3<br>CH<br>CH2<br>CH<br>CH2<br>N(CH3)<br>CH2<br>CH2<br>CH2 | CH | CH2<br>CH2 | CH2<br>CH2<br>CH2<br>CH2<br>CH<br>CH<br>CH<br>CH | CH2<br>CH2<br>CH2<br>CH2<br>CH2<br>CH<br>CH<br>CH2 | CH<br>CH<br>CH<br>CH2<br>CH<br>CH | CH<br>CH2<br>CH2<br>CH | CH2<br>CH<br>CH2<br>CH<br>CH<br>CH2<br>CH2<br>CH2<br>CH2<br>CH |
| [4.0,4.1] | [4.1,4.2] | [4.2,4.3] | [4.3,4.4] | | | | | | |
| CH2<br>CH<br>CH<br>CH | | CH2<br>CH | CH<br>CH2 | | | | | | |

Table 7. Contributions of molecular groups associated with the 35 brain metabolites.

To this end, we conducted a series of experiments. Firstly, we computed the fractal dimension of the data set object associated with the interval [0.3, 0.4]. This meant working in a space of about 20 dimensions (0.1 ppm resolution width). Figure 1 consists of the graph obtained when applying the algorithm presented in section 3.2 to the data set object associated with the interval [0.3, 0.4]. By inspection of the graph, we can observe three regions which can be differentiated by the slope of the graph. The first region is the smallest one and goes from $\log(1/r) = 0$ (hypercube size equal to 1) up to $\log(1/r) \approx 0.5$. For this interval the slope is approximately zero. The next interval goes from $\log(1/r) \approx 0.5$ up to $\log(1/r) \approx 3$ and the slope is approximately 1.6. Finally, the last interval has a slope of approximately 1. Therefore, the fractal dimension is approximately 1.6. It is important to note that we have adopted a conservative approach, instead of averaging the slopes of the graph we take the worst case behaviour of the data set object measured in terms of complexity. Therefore, the samples would be in a volume of about 2 dimensions. That is, two degrees of freedom are responsible for the observed complexity.

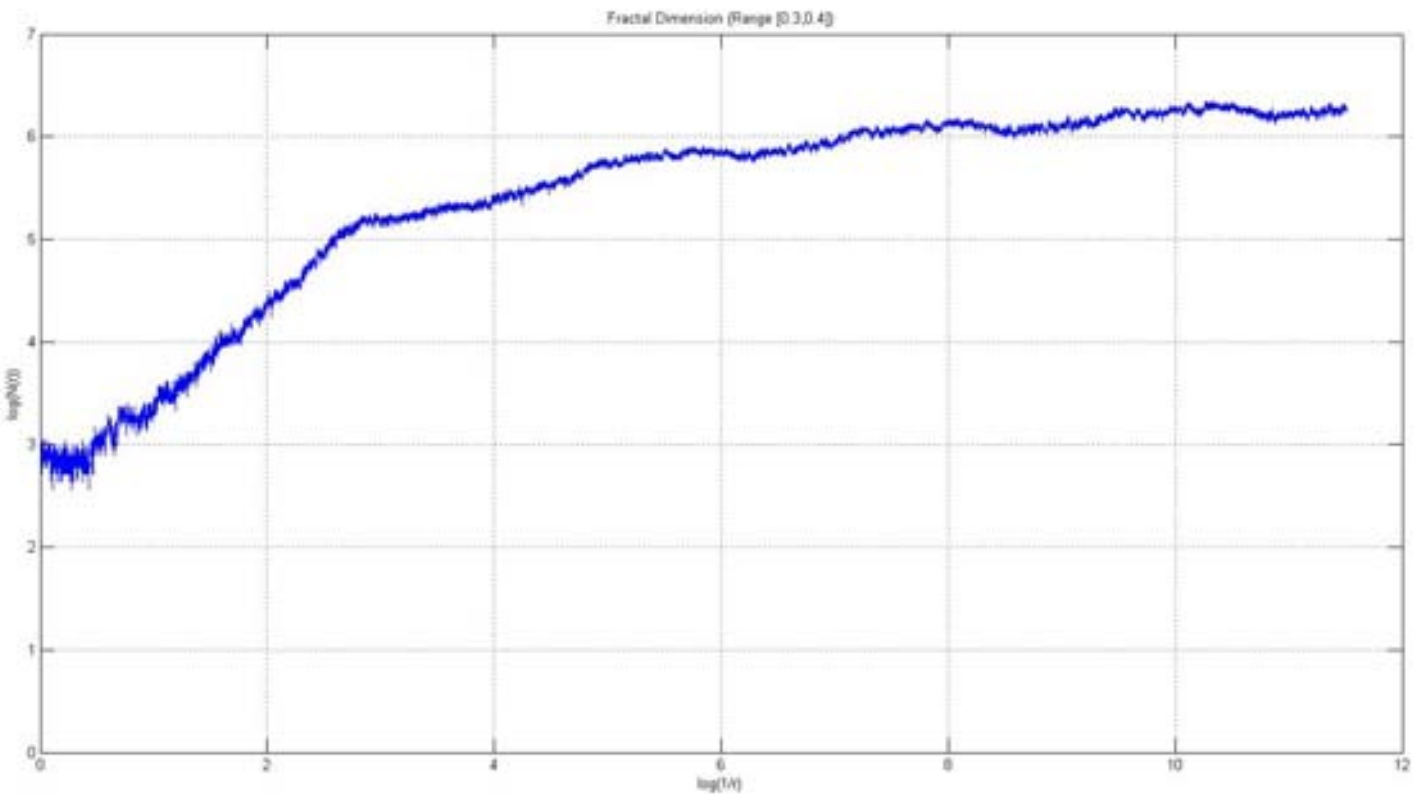

Figure 1. Fractal dimension computation for the interval [0.3, 0.4] ppm using bins of 0.1 ppm width and computed for data set A.

If we proceed in a similar way but now taking as a reference for the computation the interval [3.9, 4.0] we obtain the results shown in figure 2. A quick inspection of the graph shows an increase in complexity as expected. Specifically, there is a region where the slope of the curve is approximately 4.5. Thus, in this case the fractal dimension is approximately 4.5, which means that the samples would fit in a volume of about 5 dimensions. Furthermore, the fact that the region has the highest amount of molecular group contributions does not lead to an explosion in the complexity associated with this region but rather a slight increase in complexity - there are five degrees of freedom. At this point, it is important to emphasize the fact that the fractal dimension of a random process is infinite. Therefore, we would observe regions of the curve with an infinite slope. Additionally, although not shown here, we performed the same computation for each of the bins associated with the interval [-0.8, 4.3], and we found that complexity is bounded within the integer interval [2,5].

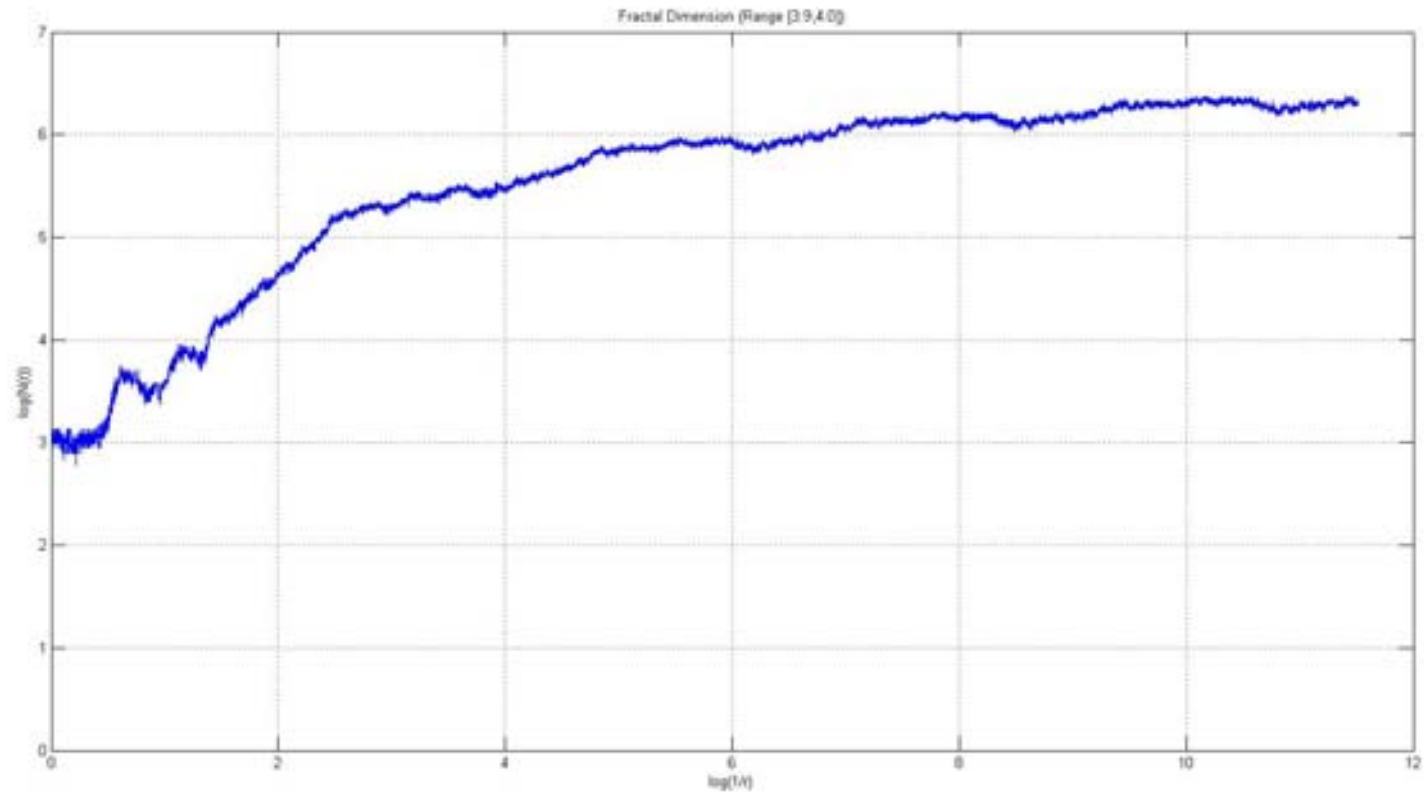

Figure 2. Fractal dimension computation for the interval [3.9, 4.0] ppm using bins of 0.1 ppm width and computed for data set A.

In turn, we performed a PCA analysis of each of the bins composing the range of interest of [-0.8, 4.3] ppm associated with brain metabolites. This analysis involved applying the PCA transformation to 52 smaller data sets. Each data set represents the data contained in a bin of 0.1 ppm resolution width constructed from the entire data set A. As a result of this procedure, the 52 data sets are composed of samples in an approximately 20-dimensional space. Furthermore, we performed the PCA transformation keeping 98% inertia. In other words, allowing only 2% information loss.

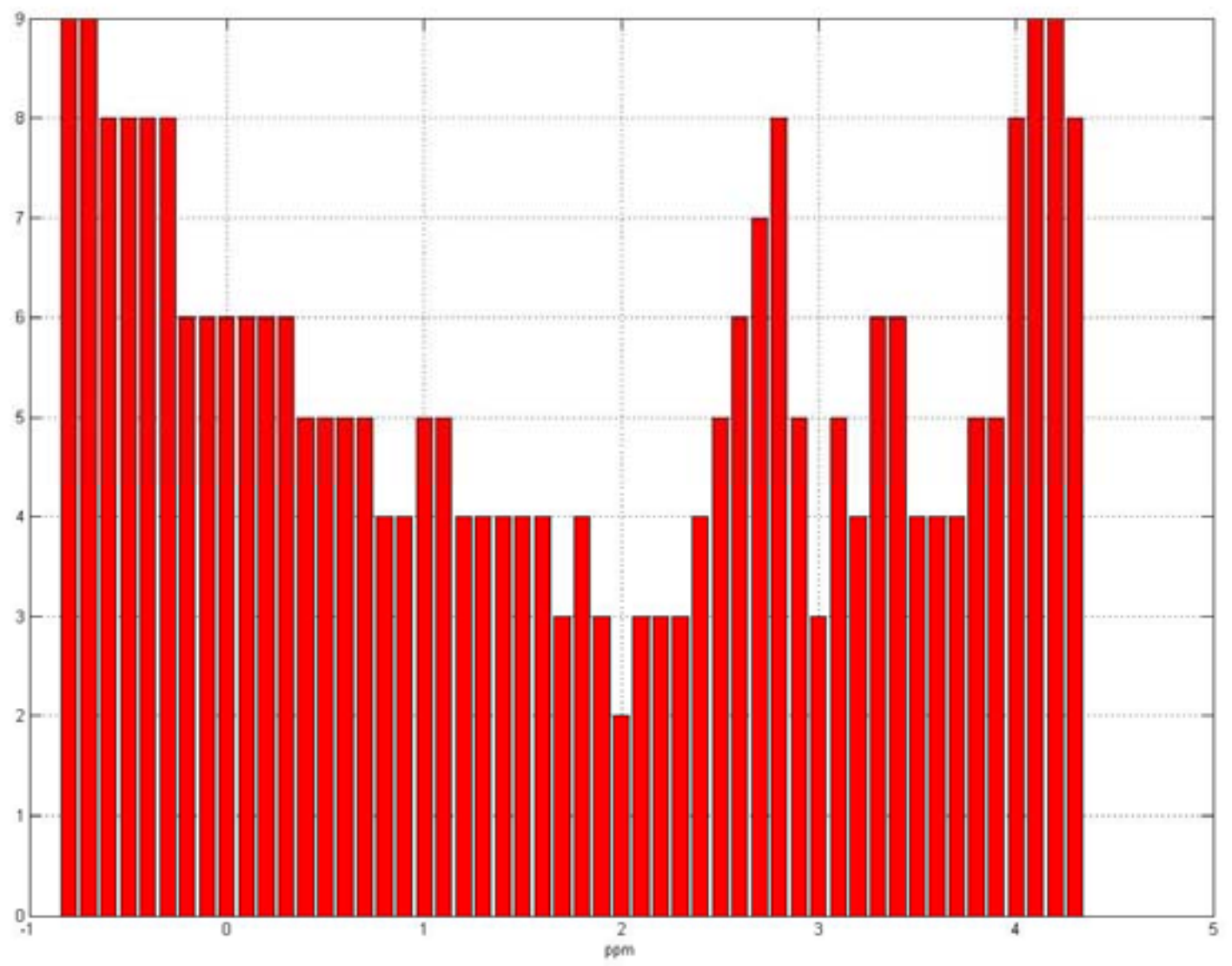

Figure 3. Principal component analysis for the interval [-0.8, 4.3] ppm using bins of 0.1 ppm width and computed from data set A.
.

Figure 3 depicts the results of the transformation. The horizontal axis represents the chemical shift scale in parts per million (ppm) while the vertical axis represents the dimensionality reduction achieved (number of dimensions needed for the above PCA parameterization). From inspection of the graph we can account for three different behaviours:

1) Regions where contributions of molecular groups are either minimal (i.e., one or two group contributions) or inexistent and the PCA needs a considerable number of dimensions (8 to 9 dimensions) to explain the variance of the data. For instance, the interval [-0.8, 0.3] and the interval [4.0, 4.4] are clear examples of such behaviour.

2) Regions where the contributions of molecular groups are minimal or inexistent but the PCA achieves a considerable reduction in dimensionality of the data set samples. The interval [0.8, 1.8] is a good example of this description.

3) Regions where there is a large amount of molecular group contributions (around 7 contributions on average) and the PCA achieves a considerable dimensionality reduction (4 to 5 dimensions). For instance, the intervals [3.5, 4.0] and [2.9, 3.2].

Taking into account the considerations stated above, it can be observed that there are regions in which the PCA needs more dimensions to explain the variance of the data than in others. Furthermore, the information is distributed differently depending on the region under consideration. In particular, concerning point three, these results suggest the existence of strong quasi-linear correlations between the metabolic signals associated with these regions, which is logical taking into account the amount of molecular group contributions associated with these regions.

However, the first and second points are to a certain degree problematic. For the first point, the results observed might suggest the existence of noise in these bands. It is important to highlight the fact that the PCA transform cannot reduce the dimensionality of a random process as it is unable to find any structure in the data. Nevertheless, this hypothesis is in contradiction with the results of complexity obtained with the fractal dimension computation. Specifically, these results indicated the presence of a system with a few degrees of freedom. As stated previously, the fractal dimension of a random process is infinite. Therefore, a plausible hypothesis for explaining these results is the existence of non-linear correlations between the metabolic signals associated with these regions. Furthermore, the PCA is limited by virtue of being a linear technique. It may therefore be unable to capture more complex non-linear correlations.

On the other hand, the regions of the spectrum considered in the second point are characterized by the absence of molecular group contributions, and hence by the apparent absence of information. Furthermore, the complexity associated with these regions is very low (i.e: 2 to 3 degrees of freedom). In addition, the PCA analysis also shows the existence of strong quasi-linear correlations. In order to explain these results it is important to note first that the absence of metabolic group contributions is not a sufficient condition to justify the presence of noise. In this case, we observe a slight discrepancy between the degrees of freedom obtained with the PCA and with the fractal dimension computation. This fact seems to justify again the hypothesis for the existence of non-linear correlations since the PCA seems to be overestimating the true dimensionality of the data sets. Indeed, the hypothesis of non-linear correlations would also fit the description of results presented for the third observation.

In order to shed some light on the previous results, we computed the entropy associated with each of the metabolic signals throughout the range [-0.8, 4.3]. The idea was to consider each metabolic signal as a realization of an unknown stochastic process. This is a natural way of addressing the positional noise common to dynamic NMR data that was described in section 2. From the definition provided in section 4.1.1 we computed the entropy associated with each of the stochastic processes that compose data matrix A. It is important to remember that each metabolic signal here is a time series composed of spectral amplitudes.

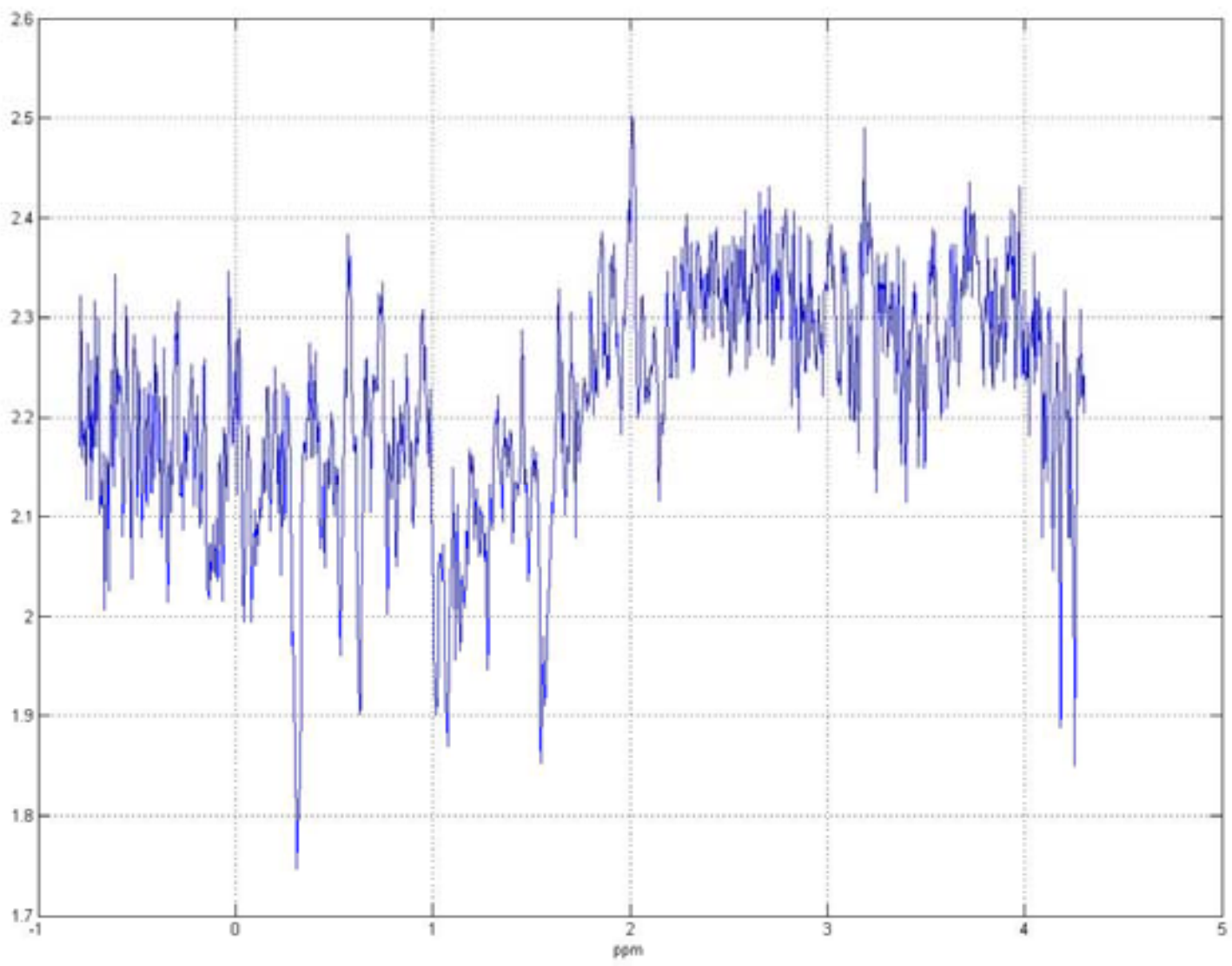

Figure 4. Entropy of the set of metabolic signals (stochastic processes) associated with the interval [-0.8, 4.3] ppm computed from data set A.

Figure 4 illustrates the results of the entropy computation. The horizontal axis represents the chemical shift scale in parts per million (ppm) while the vertical axis represents entropy in nats. From inspection of the graph, it is easy to deduce that the values of entropy present strong oscillations, even for metabolic signals whose associated resonance frequencies are contiguous or very close in the spectrum. These oscillations account for the complexity of the information distribution. Moreover, by analyzing the values of entropy reached, we can sub-divide the graph into the following logical regions:

1) The region from [2.0, 4.0] ppm is the region where average entropy is higher compared to the rest of the intervals under consideration. Furthermore, this region is characterized by the fact that it has the highest amount of group contributions (see table 7 for details). Therefore, as may be expected, the degree of randomness (i.e. disorder) or uncertainty associated with such stochastic processes is on average higher than that of other regions of the spectra. However, it is important to note that having higher entropy values also means that on average whenever we observe such processes, the amount of information conveyed by each realization of the processes is also higher. For instance, the sub-interval [3.0, 4.0] contains three local maxima which are located within the sub-intervals [3.2, 3.3], [3.7, 3.8] and [3.9, 4.0] respectively. Each of them corresponds to the ranges where the highest concentrations of molecular group contributions exist. Additionally, of particular interest is the sub-interval [2.7, 2.9], where we observe that there are several peaks corresponding to local maxima of entropy. However, from inspection of table 7, we see that the molecular group contributions are scarce, just one group contribution within the sub-interval [2.8, 2.9]. Furthermore, if we observe the results shown in figure 3, we see that the degrees of freedom or intrinsic dimension of the data found by the PCA are slightly superior to those found by the fractal dimension computation. This apparent contradiction can be partially solved if we assume again the existence of non-linear correlations, as they would explain the observed discrepancy between the degrees of freedom found by PCA and by the fractal dimension computation. However, this would not explain the unusual combination of information distribution and intrinsic dimension of data.

2) The region from [-0.8, 1.5] is the region where average entropy reaches lower values than in other regions of the spectrum. Indeed, global minimum entropy is reached within this interval. In addition, as opposed to region (1), this region is characterized by the absence of molecular group contributions. Following similar reasoning to that sketched above, whenever we observe any of the stochastic processes associated with this region, the average information conveyed per observation is lower as compared to those of region (1). Therefore, these processes (i.e. time series) are more predictable. On the other hand, it is important to note the existence of a series of maxima within this region. Specifically, there are four local maximum located within the sub-intervals [-0.7, -0.6], [-0.1, 0.0], [0.5, 0.6] and [0.7, 0.8]. The most important point to note is that they reach values which are close to the values reached in region (1) meaning that they are information-rich. However, this is in contradiction with the fact that there are no molecular group contributions in this region. Furthermore, similar to what we found in region (1), there is also a discrepancy between the degrees of freedom obtained with the PCA and the fractal dimension computation, especially for the first two peaks. Bearing in mind the fact that this region, even for short TE, is linked to effects associated with macro-molecules and lipids, a plausible hypothesis is that this information may correspond to unassigned lipid resonances. However once again such a hypothesis would not completely explain the unusual combination of information distribution, degree of randomness (uncertainty) and intrinsic dimension of data.

3) The regions [1.5, 2.0] and [4.0, 4.3] constitute a special case as they correspond to regions also characterized by the complete absence of, or the presence of very few molecular group contributions but the values of entropy reached in these regions are comparable on average to those of region (1). Concerning interval [1.5, 2.0], it is important to note that the values of entropy show an increasing tendency throughout this interval. Indeed, the global maximum entropy is reached in sub-interval [2.0, 2.1] corresponding to the N-Acetyl aspartate (NAA) resonance. This is particularly interesting as we have a region where, firstly the degree of randomness is the highest compared to the rest of the regions associated stochastic processes which are information-rich, but at the same time we have an underlying system which has few degrees of freedom (only three).

To summarize, the highest entropy values correspond in general to regions where there is a significant amount of molecular group contributions. Conversely, regions with little or no molecular group contributions are characterized by lower entropy values. As stated above, this empirical rule is only partially true as it shows certain exceptions such as the unusual relationship observed between information distribution, degree of randomness and the underlying complexity shown by the data. However, the puzzle is solved, as it is demonstrated shortly, if we think in terms of a chaotic system. A chaotic system is a deterministic system in the sense that its operation is governed by fixed rules, yet such a system with only a few degrees of freedom can exhibit behaviour so complex that it looks random. Indeed, second order statistics of a chaotic time series seem to indicate that it is a random process.

In order to visualize the nonlinear properties of the temporal NMR data, we used the Poincaré plots representation explained in section 4.1.2 to draw the phase space of metabolic signals associated with major brain metabolites. In particular, in order to simplify the problem we worked for representational purposes with uni-dimensional stochastic processes. Specifically, from the matrix associated with data set A, we drew the phase space of the metabolic signal (dimension) associated with a given metabolite having the highest entropy value. For instance, γ-Aminobutyric acid (GABA) has three methylene groups which have resonance multiplets at 1.89, 2.28 and 3.01 ppm respectively. The multiplet centered at 2.28 ppm presents the highest value of entropy so it is chosen for the phase space representation. At this point, it is important to remember again that each metabolic signal corresponds to a time series represented as a column in the matrix of data.

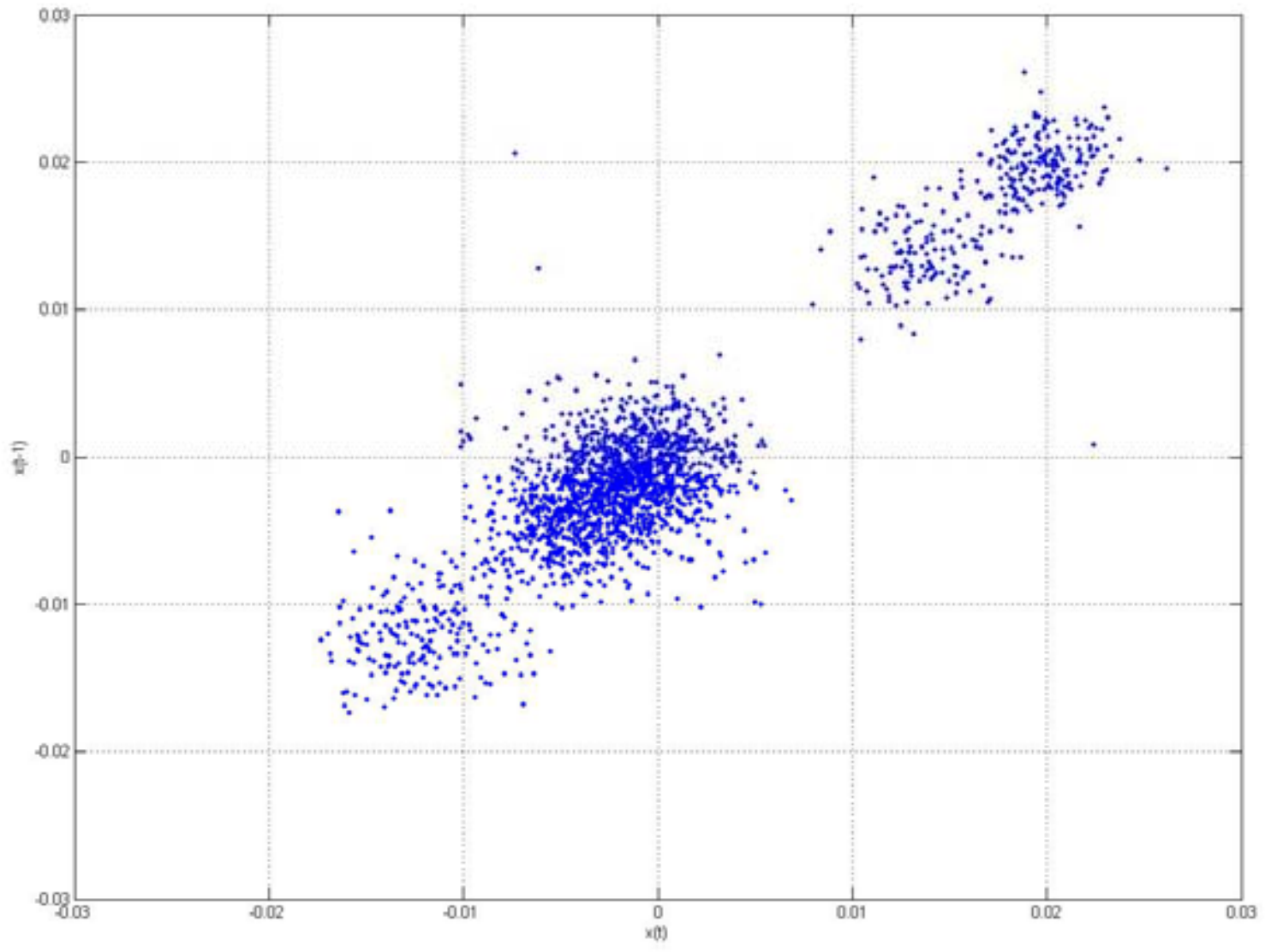

Figure 5. Phase Space representation associated with the NAAG time series.

Figures 5 and 6 represent the phase space of the time series associated with the excitatory neurotransmitters N-Acetylaspartylglutamate and glutamate, respectively, for a unitary delay. N-Acetylaspartylglutamate (NAAG hereafter) is a dipeptide of N-substituted aspartate and glutamate. It has been suggested that it is involved in excitatory neurotransmission as well as being a source of glutamate, although its function remains to be clearly established. For representation purposes we chose the singlet resonance at 2.04 ppm of the acetyl-$CH_3$ protons. In addition, glutamate is an amino acid with an acidic side chain. It is the most abundant amino acid found in the human brain. It is known to act as an excitatory neurotransmitter, although it is believed to have other functions too. It has two methylene groups and a methine group which are strongly coupled. In this case, we again opted for phase space representation, the resonance having maximum entropy value which occurs at 3.7433 ppm. The structure of the graphs seems to confirm the hypothesis that we are in the presence of a chaotic system. The most important point to note is that the processes represented are not random. In fact, they present the typical aspect of the phase space of a chaotic attractor. Random processes are characterized by a randomly distributed phase space (e.g. points uniformly distributed around the phase space) with uncorrelated data points. However, we observe here a specific structure that reflects the ergodicity of the underlying stochastic process. It is important to remember that the standard deviation of the data points perpendicular to the line of identity describes short-term variability of the stochastic process. Conversely, the standard deviation along the line of identity describes long term variability. Therefore, it is easy to deduce that the short-term and long-term variability of NAAG process are higher when compared to those of glutamate. Furthermore, its phase diagram suggests the existence of four attracting regions of phase space trajectories as compared to the two regions that can be observed in the glutamate phase space, showing that the former is apparently more complex or at least information-rich.

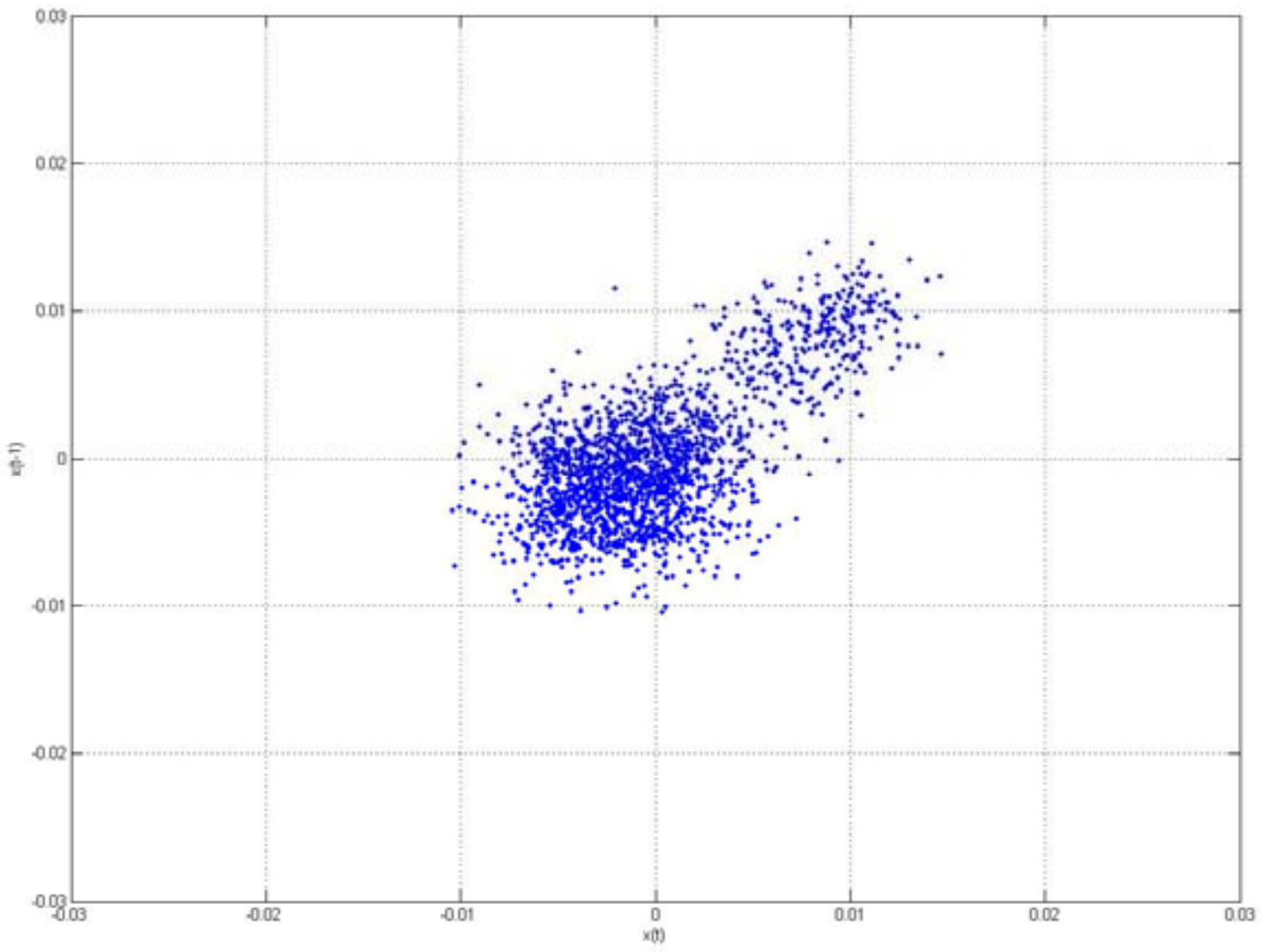

Figure 6. Phase Space representation associated with glutamate time series.

For comparison purposes we also draw the corresponding phase space associated with an inhibitory neurotransmitter like GABA and also those of of brain metabolites not involved in neurotransmission like myo-inositol and lactate. In all cases a unitary delay was used as for the previous graphs. For instance, figure 7 depicts the phase diagram of GABA. As stated above, GABA is a primary inhibitory neurotransmitter in the cerebral cortex [99]. This metabolite is synthesized from glutamate in specialized cells. The release of GABA inhibits the electrical activity of the neurons to which it is connected. At clinical field strengths all resonances from GABA overlap with other metabolite signals. As stated above, we used for representation purposes the resonance leading to an entropy maximum which occurs at 2.284 ppm. Compared to the excitatory neurotransmitters, GABA shows less marked long-term variations. In addition, in terms of attracting regions, it shows similar behaviour to that observed with glutamate, although in this case the short-term variations are less marked when compared to those observed in glutamate.

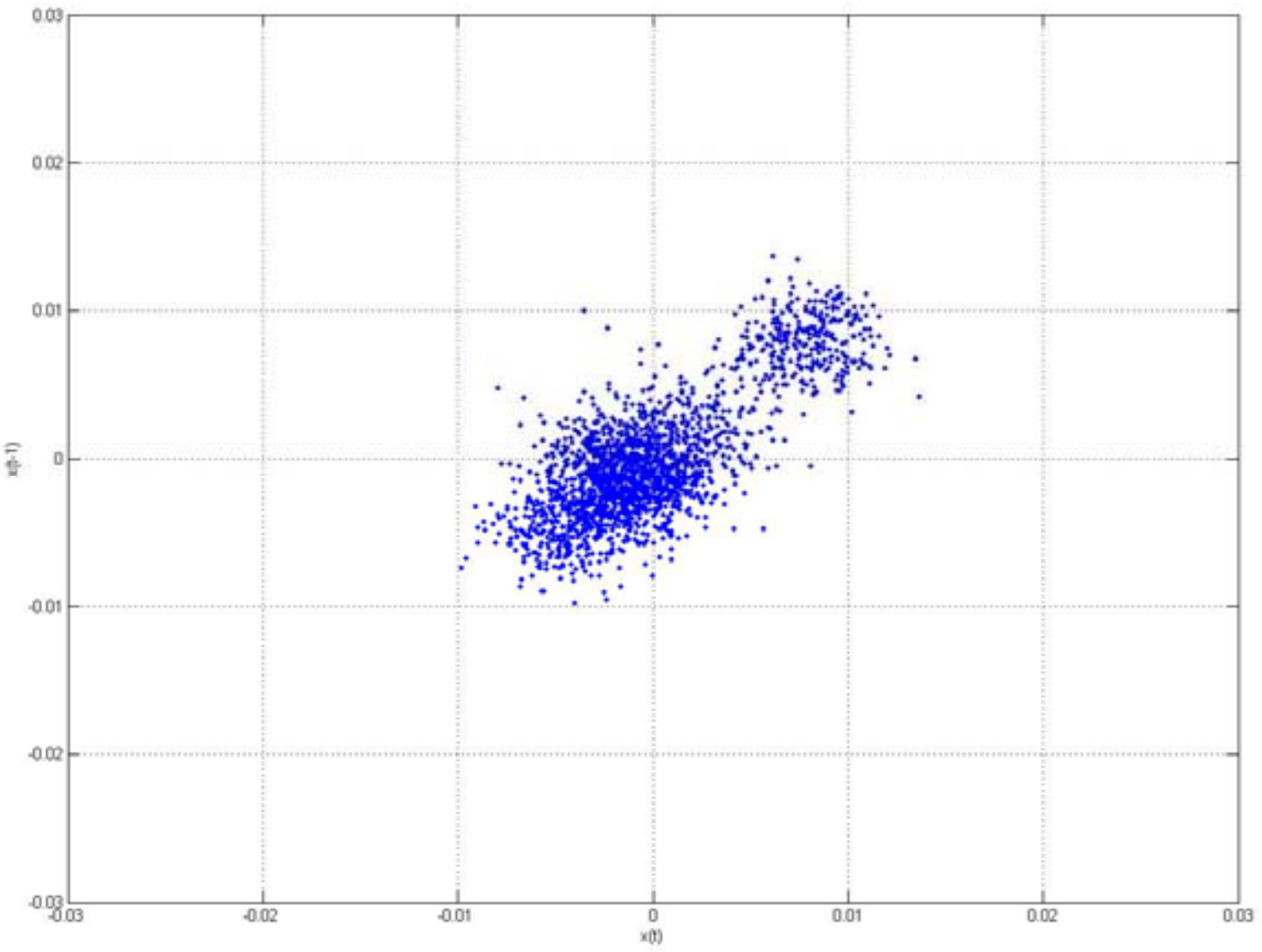

Figure 7. Phase Space representation associated with the GABA time series.

Figure 8 depicts the myo-inositol phase space diagram. Myo-inositol is a structural component of brain tissue being one of the nine isomers of inositol. The function of myo-inositol is not well understood, although it is believed to be an essential requirement for cell growth as well as a storage form of glucose. Here we used again the resonance having the highest entropy value which occurs at 3.5217 ppm. From inspection of the graph, we can see that this metabolite presents a phase diagram slightly different from those associated with neurotransmitters. In this case, we cannot distinguish different regions like we could before. However, its long-term variations are similar to those found for the GABA neurotransmitter.



Figure 8. Phase Space representation associated with myo-inositol time series.

Finally, we looked at lactate, which is a metabolically important marker of tissue ischemia and hypoxia, since its concentration is significantly increased in aerobic tissues like brain when deprived of oxygen for even short periods of time. Its detection is linked to the resonance at 1.31 ppm due to a doublet of its methyl group. However, it is difficult to reliably estimate its concentration as it belongs to a spectral region where there are many resonances from lipids. In this case, we again used the resonance having the highest entropy value which occurs at 4.0974 ppm. Figure 9 depicts its phase diagram. As can be observed, its structure, once again, is different from the structure observed for the neurotransmitters. Indeed, it is closer to myo-inositol with regard to the number of attracting regions (only one). Furthermore, the long-term variation associated with this process is lower when compared to the rest of the processes seen so far.

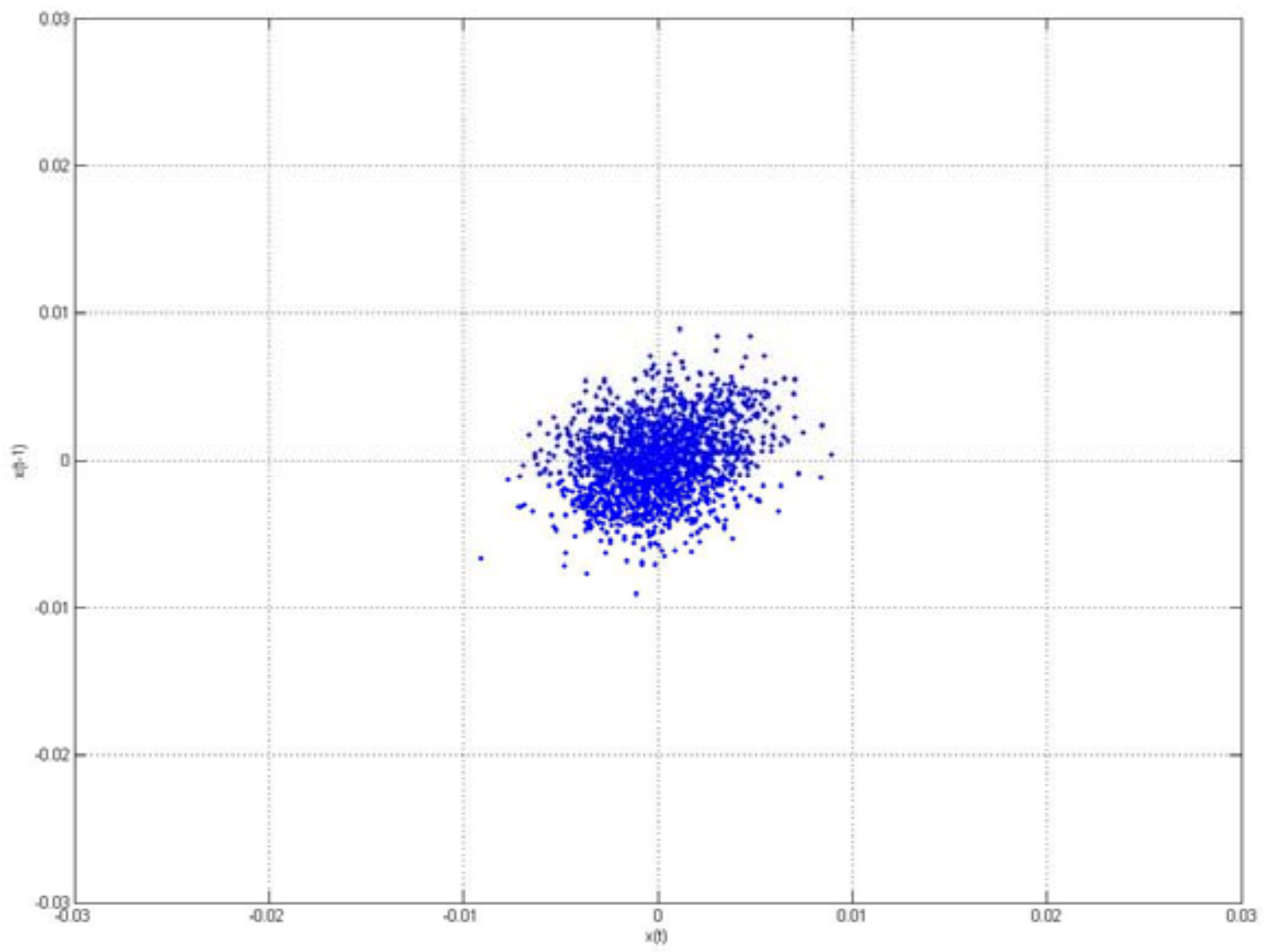

Figure 9. Phase Space representation associated with lactate time series.

On the other hand, it is important to note that we are dealing with finite amounts of data. Therefore, in order to better exploit the available information, using a unitary delay for computing the phase space diagrams can hide important information about the structure of the chaotic process under consideration. Unitary delays lead to delay vectors which are all concentrated around the diagonal (because of the correlations) in the embedding space and thus the structure perpendicular to the diagonal may not be visible. Let us denote by $y(n)$ the time series associated with any of the metabolites´ resonances. The optimal choice for the embedding delay $\tau$ is that value of $\tau$ that makes $y(n)$ and $y(n- \tau)$ independent, i.e. having no correlation with each other.

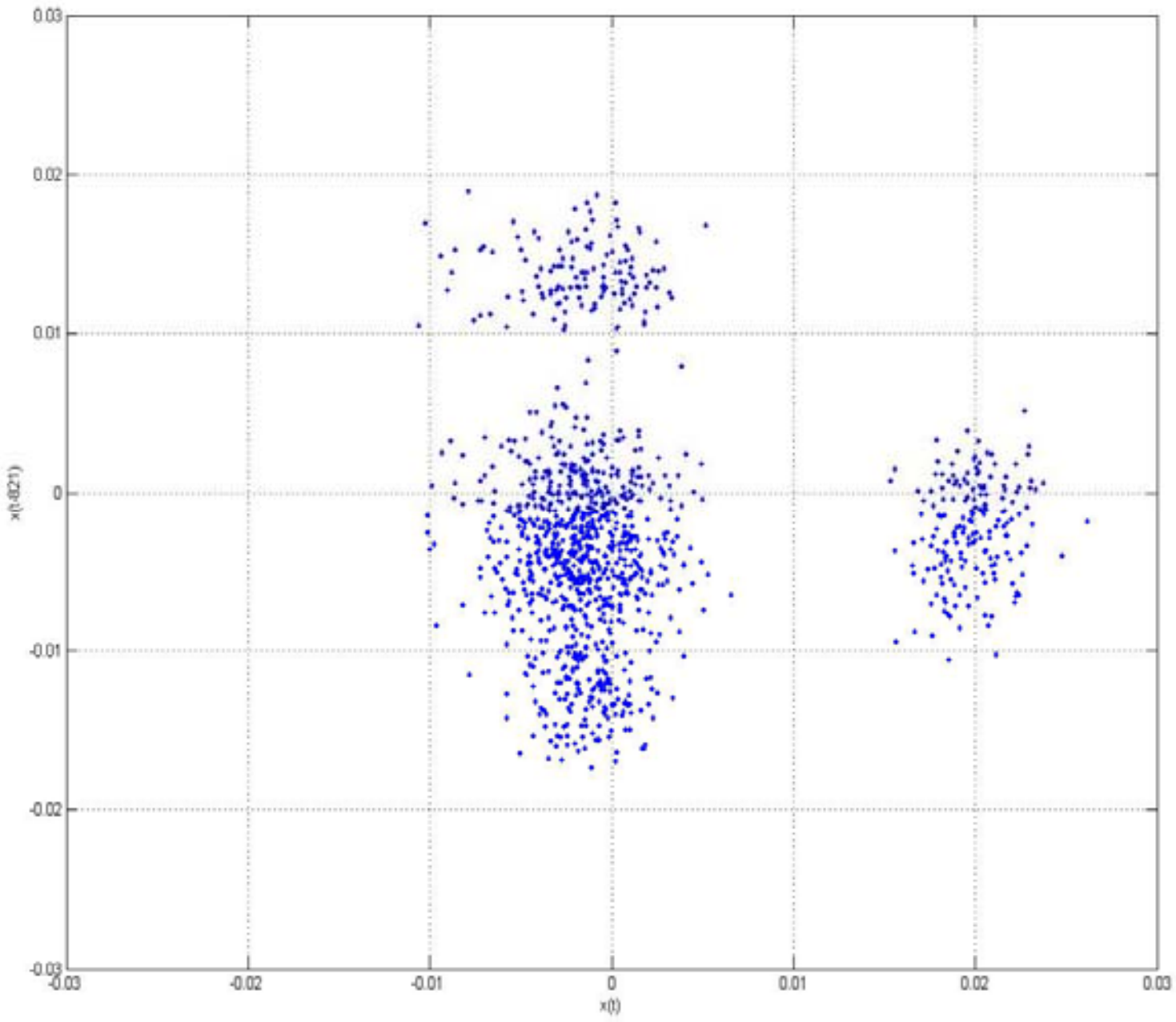

Figure 10. Phase Space representation of the NAAG time series using the optimal embedding delay computed with the mutual information procedure [100].

As stated in [100] this requirement is best satisfied by using the particular τ for which the mutual information between *y(n)* and *y(n-τ)* attains its first minimum. Taking these considerations into account we computed the optimal embedding delay for the time series associated with NAAG, GABA and myo-inositol, obtaining the embedding delay values of 821, 155 and 239 respectively.

Figure 10 displays the phase diagram of NAAG but using the delay coordinates computed with the method described above based on mutual information. As can be observed, the shape of the phase space changes substantially when using the appropriate delay coordinates. In particular, we observe three completely differentiated attracting regions. Moreover, in order to give formal proof of the chaotic behaviour we computed the maximum Lyapunov exponent of the NAAG time series using the method proposed in [96], and obtaining a value of $\lambda_{max} \approx 0.014$, thereby confirming the hypothesis of chaos. Although not shown here, similar proofs can be provided using the time series associated with the other metabolites previously studied. It is important to remember that a chaotic process is defined as a process generated by a nonlinear deterministic system with at least one positive Lyapunov exponent.

To summarize, we have proved that temporal variations of NMR metabolic signals are information-rich, by revealing their chaotic structure. Furthermore, we have also seen that the amplitude of the spectrum signal associated with a given metabolite is proportional to the concentration of that metabolite. Trajectories induced by the attractors depicted in figures 5 to 10 in fact reflect concentration changes. Furthermore, such concentration changes are not random but they follow a pattern which is specific to the metabolite under consideration. In addition, the dataset used throughout the previous analysis corresponds to NMR data coming from different individuals (collected from different brain regions) using an acquisition time big enough to observe inhibitory and excitatory neurotransmission processes. Taking these considerations into account together with the fact that the set of chemical reactions that happen in living organisms are universal, in particular metabolic pathways, it is a plausible hypothesis to state that the observed chaotic behavior inherent to metabolic signals is in fact encoding metabolic pathway information. At this point, it is important to remember that chaos is present in many human physiological processes [101] such as the cardio-respiratory system, the perception and motor control system, and voice, amongst others.

However, in order to extract such information we have to exploit the full multidimensional information associated with each metabolite and to use nonlinear filter techniques [102,105]. Nonlinear noise reduction consists of phase space reconstruction techniques that do not rely on frequency information in order to define the distinction between signal and noise. Instead, structure in the reconstructed phase space will be exploited. It is important to remember that chaotic trajectories display wide-band spectra where power decreases with the inverse of frequency. These spectra differentiate chaos from noise: white noise displays a uniform distribution of frequencies across the spectrum. Nonlinear noise reduction takes into account that nonlinear signals will form curved structures in delay space. In particular, noisy deterministic signals form blurred-out lower dimensional manifolds. Nonlinear phase space filtering tries to identify such structures and project onto them in order to reduce noise. Although further research must be carried out, these preliminary results have shown the possibility of mapping empirical NMR spectroscopy data to metabolic pathways. This fact would contribute not only to the development of early tumour detection techniques but to clarifying the high degree of information that is missing from our current understanding of complex biological systems.

## 5 Conclusions

In this paper we have investigated the application of machine learning techniques and chaos theory to characterize magnetic resonance spectroscopy data. Throughout this paper we have focused explicitly on the characterization of dynamic brain MRS data. We have presented a formal description of the problems associated with MRS-based data in terms of its application context in the field of clinical diagnosis research. Specifically, we reviewed the most common problems derived from proton scalar coupling effects, as well as those derived from differences in pH and ionic strength. Furthermore, we put especial emphasis on the fact that despite considerable progress in clinical diagnosis methodologies, current diagnosis techniques make the assumption that temporal correlations from sample to sample between metabolic signals do not carry information at all. In addition, we have introduced a machine learning framework for the analysis and characterization of MRS-based data. The principal advantage of the proposed methodology is that it is able to cope with both static and dynamic aspects of NMR-based data. The combination of a pre-processing technique which is able to exploit the correlations between metabolic signals, and a set of structural parameters suited to the high-dimensional characteristics of spectroscopy patterns allowed us to extract the relevant information required for the detection and diagnosis of disease. Moreover, in order to understand the underlying structure of dynamic NMR-based data we have also performed a detailed analysis based on information theoretic concepts and chaos theory. The results revealed an information-rich structure as shown by the chaotic nature of dynamic MRS data. Furthermore, we argued that they actually encode metabolic pathway information.

Summarizing, the main contribution of the present study is to invalidate the traditional view that disregards the dynamic aspects of MRS data as being devoid of information. Traditional methods are characterized by working with averaged data; conversely, the nonlinear analysis shown in this work attempts to identify properties of the underlying dynamics. Moreover, we can conclude, firstly that a deep understanding of the complex relationships between metabolites under normal and pathological conditions cannot be conceived without taking into account their dynamical interactions. Furthermore, the framework presented here opens the possibility of developing methods not only as a starting point for reliable diagnosis systems development but for efficiently modeling the dynamics of deep molecular pathways with high levels of noise and relatively small experimental data sizes.

Finally, it is important to note, that from our point of view the application of chaos theory to physiological data is yet widely unexplored. Therefore, a successful approach cannot be conceived without a close collaboration between individuals of different disciplines. We hope to have provided sufficient motivation for further studies and applications as we believe it is a great challenge to adopt these methods and to apply them in the clinical research field.

**Acknowledgements.**

The author acknowledges Dr. J.L. González Mora for providing the NMR data used in this work.